\documentclass[3p,times]{elsarticle}

\usepackage{ecrc}

\volume{00}

\firstpage{1}

\journalname{Physica A}

\runauth{P.~Jizba and J.~Korbel}

\jnltitlelogo{Physica A}

\CopyrightLine{2014}{Published by Elsevier Ltd.}

\usepackage{amssymb}
\usepackage{amsthm}
\usepackage[usenames,dvipsnames]{color}
\usepackage{amsmath}
\usepackage{dsfont}
\usepackage{graphicx}
\usepackage{algorithm}
\usepackage[usenames,dvipsnames]{color}
\usepackage{amsmath}
\usepackage{dsfont}
\usepackage{graphicx}
\newcommand{\ud}{\mathrm{d}}

\definecolor{darkred}{rgb}{.8,0,0}

\definecolor{darkblue}{rgb}{0,0,.7}

\def\noi{\noindent}

\begin{document}

\begin{frontmatter}

%% Title, authors and addresses

%% use the tnoteref command within \title for footnotes;
%% use the tnotetext command for the associated footnote;
%% use the fnref command within \author or \address for footnotes;
%% use the fntext command for the associated footnote;
%% use the corref command within \author for corresponding author footnotes;
%% use the cortext command for the associated footnote;
%% use the ead command for the email address,
%% and the form \ead[url] for the home page:
%%
%% \title{Title\tnoteref{label1}}
%% \tnotetext[label1]{}
%% \author{Name\corref{cor1}\fnref{label2}}
%% \ead{email address}
%% \ead[url]{home page}
%% \fntext[label2]{}
%% \cortext[cor1]{}
%% \address{Address\fnref{label3}}
%% \fntext[label3]{}

\dochead{}
%% Use \dochead if there is an article header, e.g. \dochead{Short communication}
%% \dochead can also be used to include a conference title, if directed by the editors
%% e.g. \dochead{17th International Conference on Dynamical Processes in Excited States of
%% Solids}

%%%%%%%%%%%%%%%%%%%%%%%%%%%%%%%%%%%%%%%%%%%%%%%%%%%%%%%%%%%%%%%%%%%%%%%
\title{On $q$-non-extensive statistics with non-Tsallisian entropy}
%%%%%%%%%%%%%%%%%%%%%%%%%%%%%%%%%%%%%%%%%%%%%%%%%%%%%%%%%%%%%%%%%%%%%

%% use optional labels to link authors explicitly to addresses:
%% \author[label1,label2]{<author name>}
%% \address[label1]{<address>}
%% \address[label2]{<address>}

\author[FNSPE,FU]{Petr Jizba}
\ead{p.jizba@fjfi.cvut.cz}

\author[FNSPE]{Jan Korbel}
\ead{korbeja2@fjfi.cvut.cz}

\address[FNSPE]{Faculty of Nuclear Sciences and Physical Engineering, Czech Technical University in Prague, B\v{r}ehov\'{a} 7,
11519, Prague, Czech Republic}
\address[FU]{Institute of Theoretical Physics, Freie Universit\"{a}t in Berlin, Arnimallee 14, 14195 Berlin, Germany}

\begin{abstract}
We combine an axiomatics of R\'{e}nyi with the $q$--deformed
version of Khinchin axioms to obtain a measure of information
(i.e., entropy) which accounts both for systems with embedded
self-similarity and non-extensivity.
%Such measures could be
%particularly pertinent for e.g., critical phenomena or in a strong
%quantum regime.
%We analyze
%the axiomatic foundations of R\'{e}nyi and
%Tsallis--Havrda--Charvat entropies. We show that the key elements
%of respective axiomatics (i.e., quasi--linear averaging and
%$q$--additivity) can be naturally embodied in yet new axiomatic
%prescription.
We show that the entropy thus obtained is uniquely solved in terms
of a one-parameter family of information measures. The ensuing
maximal-entropy distribution is phrased in terms of a special function
known as the Lambert W--function. We analyze the
corresponding ``high" and ``low-temperature" asymptotics and
reveal a non-trivial structure of the parameter space.
%and graphically represented.
%
%
\end{abstract}

\begin{keyword}
%% keywords here, in the form: keyword \sep keyword
Multifractals \sep R{\'e}nyi's information entropy \sep THC
entropy \sep MaxEnt \sep Heavy-tailed distributions
%% PACS codes here, in the form: \PACS code \sep code
\PACS 65.40.Gr, 47.53.+n, 05.90.+m
%% MSC codes here, in the form: \MSC code \sep code
%% or \MSC[2008] code \sep code (2000 is the default)
\end{keyword}

\end{frontmatter}

%%%%%%%%%%%%%%%%%%%%%%%%%%%%%%%%%%%%%%%%
\section{Introduction}\label{I}
%%%%%%%%%%%%%%%%%%%%%%%%%%%%%%%%%%%%%%%%

In his 1948 paper~\cite{Shannon1} Shannon formulated the theory of
data compression. The paper established a fundamental limit to
lossless data compression and showed that this limit coincides
with the information measure presently known as Shannon's entropy
${\mathcal{H}}$.
%The exact value of ${\mathcal{H}}$ depends on the
%information source
%--- more specifically, the statistical nature of the source.
In words, it is possible to compress the source, in a lossless
manner, with compression rate close to ${\mathcal{H}}$, it is
mathematically impossible to do better than ${\mathcal{H}}$.
However, many modern communication processes, including signals,
images and coding/decoding systems, often operate in complex
environments dominated by conditions that do not match the basic
tenets of Shannon's communication theory. For instance, buffer
memory (or storage capacity) of a transmitting channel is often
finite, coding can have a non--trivial cost function, codes might
have variable-length codes, sources and channels may exhibit
memory or losses, etc. Information theory offers various
generalized (non--Shannonian) measures of information to deal with
such cases.
%The
%latter substitutes Shannon's information measure with information
%measures of other types.
Among the most frequently used one can mention, e.g., Havrda--Charv\'{a}t
measure~\cite{HaCh}, Sharma--Mittal measure~\cite{SM1}, R\'{e}nyi's
measure~\cite{Re2} or Kapur's measures~\cite{Kap1}. Information
entropies get even more complex by considering communication
systems with quantum channels~\cite{Lennert,Coles}. There exists even attempts
to generalize Shannon's measure of information in the direction
where no use of the concept of probability is needed hence
demonstrating that information is more primitive notion than
probability~\cite{Urbanik1}.

In mid 1950 Jaynes~\cite{Jaynes57} proposed the {\em Maximum
Entropy Principle} (MaxEnt) as a general inference procedure that,
among others, bears a direct relevance to statistical mechanics
and thermodynamics. The conceptual frame of Jaynes's MaxEnt is
formed by Shannon's communication theory with Shannon's
information measure as an inference functional. The central
r\^{o}le of Shannon's entropy as a tool for inductive inference
(i.e., inference where new information is given in terms of
expected values) was further demonstrated in works of
Faddeyev~\cite{Fad1}, Shore and Johnson~\cite{SJ1},
Wallis~\cite{Wal1}, Tops\o{e}~\cite{Top1} and others.
%It should be noted various historical precursors in statistical
%physics can be traced already to Elsasser~\cite{Elsasser} and
%notably to work of Gibbs~\cite{}.
In Jaynes's procedure the laws of statistical mechanics can be
viewed as {\em inferences} based entirely on prior information
that is given in the form of expected values of energy, energy and
number of particles, energy and volume, energy and angular
momentum, etc., thus re-deriving the familiar canonical ensemble,
grand-canonical ensemble, pressure ensemble, rotational ensemble,
etc., respectively~\cite{Jaynes55}. Remarkable feature of this
procedure is that it entirely dispenses with such traditional hypotheses as
ergodicity or metric transitivity. Following
Jaynes, one should view the MaxEnt distribution (or maximizer) as
a distribution that is maximally noncommittal with regard to
missing information and that agrees with all what is known about
prior information, but expresses maximum uncertainty with respect
to all other matters~\cite{Jaynes57}. By identifying the
statistical sample space with the set of all (coarse-grained)
microstates the corresponding maximizer yields the Shannon entropy
that corresponds to the Gibbs entropy of statistical physics.

Surprisingly, despite the aforementioned connection between information theory and physics
and despite related advancements in non-Shannonian information theory, tendencies aiming at
similar extensions of the Gibbs's entropy paradigm started to penetrate into
statistical physics only in the last two decades. This happened
when evidence accumulated showing that there are indeed many
situations of practical interest requiring more ``exotic"
statistics which do not conform with Gibbsian exponential
maximizers. Percolation, protein folding, critical phenomena,
cosmic rays, turbulence, granular matter or stock market returns might
provide examples.

In attacking the problem of generalization of Gibbs's entropy the
information theoretic route to equilibrium statistical physics
provides a very useful conceptual guide. The natural strategy that
fits this framework would be then to revisit the axiomatic rules
governing Shannon's information measure and potential extensions
translate into language of statistical physics. In fact, the
usual axiomatics of Khinchin~\cite{Kh1} is prone to several
``plausible" generalizations. Among those, the additivity of
independent mean information is a natural axiom to attack. Along
those lines, two fundamentally distinct generalization schemes
have been pursued in the literature; one redefining the
statistical mean and another generalizing the additivity rule.

The first mentioned generalization was realized by R\'{e}nyi by
employing the most general means still compatible with Kolmogorov
axioms of probability theory. These, so called, quasi-linear means
were independently studied by Kolmogorov~\cite{Ko1} and
Nagumo~\cite{Na1}. It was shown that the generalization based on
quasi-linear means unambiguously leads to information measures
known as R\'{e}nyi entropies~\cite{Re2,Re1}. Although,  the status
of R\'{e}nyi entropies (RE's) in statistical physics is still
debated, they nevertheless provide an immensely important analyzing
tool in classical statistical systems with a non-standard scaling
behavior (e.g., fractals, multifractals, etc.)~\cite{TCH1,Harte}.

On the other hand, the second approach generalizes the additivity
prescription but keeps the usual linear mean. Currently popular
generalization is the $q$-additivity prescription and related $q$-calculus~\cite{VD2,MJ2}.
The corresponding axiomatics~\cite{Ab2}
provides the entropy known as Tsallis--Havrda--Charv\'{a}t's (THC)
entropy\footnote{Other important approaches such as
Kaniadakis's~\cite{Kan1} and Naudts's~\cite{Nau1} deformed
Hartley's logarithmic information also utilize linear means and
generalized additivity rule (e.g., $\kappa$-additivity) but as yet
they still lack the information-theoretic axiomatics that is
crucial in our reasonings. For this reason we exclude these works
from our consideration.}. As the classical additivity of
independent information is destroyed in this case, a new more
exotic physical mechanisms must be sought to comply with THC
predictions. Recent theoretical advances in systems with long-range
interactions~\cite{Ruffo}, in generalized (and specifically $q$-generalised) central limit
theorems~\cite{Suyari}, in theory of asymptotic scaling~\cite{Thurner}, etc.,
indicate that the typical playground
for THC entropy should be in cases where two statistically
independent systems have non-vanishing long-range/time
correlations or where the notion of statistical independence is an
ill-defined concept. Examples include, long-range Ising models,
gravitational systems, statistical systems with quantum
non-locality, etc.

It is clear that an appropriate combination of the above generalizations
could provide a new conceptual paradigm suitable for a
statistical description of systems possessing both
self-similarity and non-locality. Such systems are quite
pertinent with examples spanning from the early universe
cosmological phase transitions to currently much studied quantum
phase transitions (frustrated spin systems, Fermi liquids, etc.).
In passing we should mention that there exists a number of works trying to compare
both R\'{e}nyi and THC entropies   from both the theoretical and  observational point of view
(see, e.g, Refs.~\cite{Bercher,Lima}). Nevertheless,
the merger of both entropic paradigms has not been studied yet.
It is aim of this paper to pursue this line of reasonings and
explore the resulting implications.
In order to set a stage for
our considerations we review in the following section some axiomatic
essentials for both Shannon, R\'{e}nyi and THC entropies that will
be needed in the main body of the paper. In Section~\ref{VII} we then
formulate a new axiomatics which aims at bridging the R\'{e}nyi
and THC entropies. It is found that such axiomatics allows for
only one one-parametric family of solutions. Basic properties of
the new entropy that we denote as ${\mathcal{D}}_q$ are discussed.
A simplification that
${\mathcal{D}}_q$ undergoes in multifractal systems is particularly emphasized.
The corresponding MaxEnt distributions are calculated in Section~\ref{X}.
We utilize both linear and non--linear moment constraints
(applied to the energy) to achieve this goal. In both
aforementioned cases the distributions are expressible through the
Lambert W--function. Since the analytic structure of MaxEnt distributions is too complex
we confine our analysis to the corresponding ``high" and
``low-temperature" asymptotics and discuss the ensuing non-trivial
structure of the parameter space. In Section~\ref{sec: concavity} we
discuss the concavity and Schur-concavity of ${\mathcal{D}}_q$.
Section~\ref{IX} is devoted to
conclusions. The paper is substituted with three appendices which
clarify some finer mathematical points.

%%%%%%%%%%%%%%%%%%%%%%%%%%%%%%%%%%%%%%%%%%%%%%%%%%%%%%%%%%%%
\section{Brief review of entropy axiomatics}
%%%%%%%%%%%%%%%%%%%%%%%%%%%%%%%%%%%%%%%%%%%%%%%%%%%%%%%%%%%%

The information measure, or simply entropy, is supposed to
represent the measure or degree of uncertainty or expectation in
conveyed information which is going to be removed by the
recipient. As a rule in information theory the exact value of
entropy depends only on the information source
--- more specifically, on the statistical nature of the source.
Generally speaking, the higher is the information measure the
higher is the ignorance about the system (source) and thus more
information will be uncovered after the message is received (or an
actual measurement is performed). As often happens, this simple
scenario is not frequently tenable as various restrictive factors
are present in realistic situations; finite buffer
capacity, global patterns in messages, topologically non--trivial
sample spaces, etc.. One may even entertain various information
theoretic implications related with the quantum probability
calculus or quantum communication channels. Thus, as we go to
somewhat more elaborate and realistic models, the entropy
prescriptions get more complicated and realistic!

To see why a new generalization of the entropy is desirable let us
briefly dwell into 3 most common entropy protagonist, namely Shannon's,
R\'{e}nyi's and THC entropy.

%%%%%%%%%%%%%%%%%%%%%%%%%%%%%%%%%%%%%%%%%%%%%%%%%%%%%%%%%%%%
\subsection{Shannon's entropy ---  Khinchin axioms}\label{II}
%%%%%%%%%%%%%%%%%%%%%%%%%%%%%%%%%%%%%%%%%%%%%%%%%%%%%%%%%%%%

The best known and widely used information measure is Shannon's
entropy. For the completeness sake we now briefly recapitulate the
Khinchin axiomatics as this will prove important in what follows.
It consist of four axioms~\cite{Kh1}:
\begin{enumerate}
\item For a given integer $n$ and given ${\mathcal{P}} = \{ p_1,
p_2, \ldots , p_n\}$ ($p_k \geq 0, \sum_k^n p_k =1$),
${\mathcal{H}}({\mathcal{P}})$ is a continuous with respect to all
its arguments.

\item  For a given integer $n$, ${\mathcal{H}}(p_1, p_2, \ldots ,
p_n)$ takes its largest value for $p_k = 1/n$ ($k=1,2, \ldots, n$).

%${\mathcal{I}}_{\alpha}({\mathcal{P}})$ takes its largest value for
%$p_k = 1/n, (k = 1,2, \ldots, n)$, i.e., the gained information is
%largest when we known least about the original system.

\item For a given $q\in {\mathbb{R}}$; ${\mathcal{H}}(A\cup B) =
{\mathcal{H}}(A) + {\mathcal{H}}(B|A)$ with
 ${\mathcal{H}}(B|A) =
\sum_k p_k \ {\mathcal{H}}(B|A=A_k)$, and distribution
${\mathcal{P}}$ corresponds to the experiment $A$.

%For independent events, i.e., ${\mathcal{R}} = {\mathcal{P}}\times
%{\mathcal{Q}}=
%\{p_iq_k \}$: \\
%${\mathcal{I}}_{\alpha}({\mathcal{P}}\times {\mathcal{Q}}) =
%{\mathcal{I}}_{\alpha}({\mathcal{P}}) + {\mathcal{I}}_{\alpha}({\mathcal{Q}})$.
%
%\item In general case there is a continuous invertible function
%$f(x)$ such that: ${\mathcal{I}}_{\alpha}\partial
%f({\mathcal{I}}_{\alpha})/\partial
%{\mathcal{I}}_{\alpha} = g({\mathcal{I}}_{\alpha}) f({\mathcal{I}}_{\alpha})$.\\
%\\
%Here ${\mathcal{I}}_{\alpha}({\mathcal{P}}\cup {\mathcal{Q}}) =
%{\mathcal{I}}_{\alpha}({\mathcal{I}}_{\alpha}({\mathcal{P}}),
%{\mathcal{I}}_{\alpha}(\mathcal{Q}))$. The scaling function $g(x)$ is
%common to both ${\mathcal{I}}_{\alpha}({\mathcal{P}}\cup {\mathcal{Q}}),
%{\mathcal{I}}_{\alpha}({\mathcal{P}})$ and ${\mathcal{I}}_{\alpha}({\mathcal{Q}})$
%.

\item ${\mathcal{H}}(p_1,p_2, \ldots , p_n, 0 ) =
{\mathcal{H}}(p_1,p_2, \ldots , p_n)$, i.e., adding an event of
probability zero (impossible event) we do not gain any new
information.
\end{enumerate}
\noindent The corresponding information measure, Shannon's
entropy, then reads (up to the normalization constant\footnote{The normalization influences the base of the logarithm. In information theory, it is common to choose normalization  
${\mathcal{H}} ( \frac{1}{2},\frac{1}{2} ) =1$, leading to binary logarithms. We adopt physical conventions and in the whole text use the normalization leading to natural logarithms.}
\begin{eqnarray}
{\mathcal{H}}({\mathcal{P}}) = - \sum_{k=1}^n p_k \ln p_k \, .
\end{eqnarray}
\noindent In passing we should stress two important points.
Firstly, 3rd axiom (known as separability or strong additivity axiom)
indicates that Shannon's entropy of two
independent experiments (sources) is additive. Secondly,
there is an intimate connection between the Boltzmann--Gibbs entropy and Shannon's entropy.
In fact, thermodynamics can be viewed as a specific application of Shannon's information theory:
the thermodynamic entropy may be interpreted (when rescaled to ``bit'' units) as the amount of
Shannon information needed to define the detailed microscopic state of
the system, which remains ``uncommunicated'' by a description that is solely in terms of
thermodynamic state variables.

%
% by
% identifying thermodynamic microstates with all possible outcomes
% of some experiment/measurement then Shannon's entropy will
% coincide with Gibbs entropy.

%%%%%%%%%%%%%%%%%%%%%%%%%%%%%%%%%%%%%%%%%%%%%%%%%%%%%%%%%%%%%%%
\subsection{R\'{e}ny's entropy: entropy of multifractal systems
}\label{III}
%%%%%%%%%%%%%%%%%%%%%%%%%%%%%%%%%%%%%%%%%%%%%%%%%%%%%%%%%%%%%%%%

As already mentioned, RE represents a step further towards more
realistic situations encountered in information theory. Among a
myriad of information measures, RE's distinguish themselves by firm
operational characterizations. These were established by
Arikan~\cite{Arikan1} for the theory of guessing, by
Jelinek~\cite{Jelinek1} for the buffer overflow problem in lossless
source coding, by Cambell~\cite{Cambell} for the lossless
variable--length coding problem with an exponential cost
constraint, etc. Recently, an interesting operational
characterization of RE was provided by Csisz{\'a}r~\cite{Csi1} in
terms of block coding and hypotheses testing. In the latter case
the R\'{e}nyi parameter $q$ was directly related to so--called
$\beta$-cutoff rates~\cite{Csi1}.

Apart from information theory RE's have proved to be  an indispensable
tool also in numerous branches of physics. Typical examples are
provided by chaotic dynamical systems and multifractal statistical
systems (see e.g.,~\cite{PJ1} and citations therein). Fully developed
turbulence, earthquake analysis and generalized dimensions of
strange attractors provide examples.

RE of order $q$ ($q >0$) of a discrete distribution ${\mathcal{P}} =
\{ p_1, \ldots, p_n\} $ are defined as
\begin{eqnarray}
{\mathcal{I}}_q({\mathcal{P}}) = \frac{1}{(1-q)} \ln\left(
\sum_{k=1}^n (p_k)^q  \right)\, . \label{III1}
\end{eqnarray}
In his original work, R\'{e}nyi~\cite{Re2,Re1} introduced a
one-parameter family of information measures (\ref{III1}) which he
based on axiomatic considerations. In the course of time these
axioms have been sharpened by Dar\'{o}tzy~\cite{Dar1} and
others~\cite{Oth2}. Most recently it was shown that RE can be uniquely derived from the following
set of axioms~\cite{PJ1}:
\begin{enumerate}
\item For a given integer $n$ and given ${\mathcal{P}} = \{ p_1,
p_2, \ldots , p_n\}$ ($p_k \geq 0, \sum_k^n p_k =1$),
${\mathcal{I}}({\mathcal{P}})$ is a continuous with respect to all
its arguments.

\item For a given integer $n$, ${\mathcal{I}}(p_1, p_2, \ldots ,
p_n)$ takes its largest value for $p_k = 1/n$ ($k=1,2, \ldots, n$).

%${\mathcal{I}}_{\alpha}({\mathcal{P}})$ takes its largest value for
%$p_k = 1/n, (k = 1,2, \ldots, n)$, i.e., the gained information is
%largest when we known least about the original system.

\item For a given $q\in {{\mathbb{R}}}$; ${\mathcal{I}}(A\cup B) =
{\mathcal{I}}(A) + {\mathcal{I}}(B|A)$ with ${\mathcal{I}}(B|A) =
\mbox{{\textsl{g}}}^{-1} \left(\sum_k \varrho_k(q)
\mbox{{\textsl{g}}}({\mathcal{I}}(B|A=A_k)) \right)$, and
$\varrho_k(q) = p_k^{q}/\sum_k p_k^{q}$\, (distribution
${\mathcal{P}}$ corresponds to the experiment $A$).
%
%For independent events, i.e., ${\mathcal{R}} = {\mathcal{P}}\times
%{\mathcal{Q}}=
%\{p_iq_k \}$: \\
%${\mathcal{I}}_{\alpha}({\mathcal{P}}\times {\mathcal{Q}}) =
%{\mathcal{I}}_{\alpha}({\mathcal{P}}) + {\mathcal{I}}_{\alpha}({\mathcal{Q}})$.
%
%\item In general case there is a continuous invertible function
%$f(x)$ such that: ${\mathcal{I}}_{\alpha}\partial
%f({\mathcal{I}}_{\alpha})/\partial
%{\mathcal{I}}_{\alpha} = g({\mathcal{I}}_{\alpha}) f({\mathcal{I}}_{\alpha})$.\\
%\\
%Here ${\mathcal{I}}_{\alpha}({\mathcal{P}}\cup {\mathcal{Q}}) =
%{\mathcal{I}}_{\alpha}({\mathcal{I}}_{\alpha}({\mathcal{P}}),
%{\mathcal{I}}_{\alpha}(\mathcal{Q}))$. The scaling function $g(x)$ is
%common to both ${\mathcal{I}}_{\alpha}({\mathcal{P}}\cup {\mathcal{Q}}),
%{\mathcal{I}}_{\alpha}({\mathcal{P}})$ and ${\mathcal{I}}_{\alpha}({\mathcal{Q}})$
%.
Here $\mbox{{\textsl{g}}}$ is invertible and positive in $[0,
\infty)$.

\item ${\mathcal{I}}(p_1,p_2, \ldots , p_n, 0 ) =
{\mathcal{I}}(p_1,p_2, \ldots , p_n)$.
\end{enumerate}
\noindent Former axioms markedly differ from those utilized
in~\cite{Re2,Re1,Dar1,Oth2}. Particularly distinctive is the
presence of the escort (or zooming) distribution $\varrho(q)$ in
the 3rd axiom. Distribution $\varrho(q)$ was originally introduced by
R\'{e}nyi~\cite{Re2} to define the entropy associated with the
joint distribution. Quite independently was $\varrho(q)$
introduced by Beck and Schl\"{o}gl~\cite{BS1} in the context of
non-linear dynamics.

We briefly remind some elementary properties of ${\mathcal{I}}_q$: it is symmetric in all arguments,
for ${q}\leq 1$ is ${\mathcal{I}}_q$ a concave function and ${\mathcal{H}}({\mathcal{P}}) \ \leq \
{\mathcal{I}}_q({\mathcal{P}})$, while for ${q}\geq 1$ it is neither concave nor convex and ${\mathcal{I}}_q({\mathcal{P}})
\ \leq \ {\mathcal{H}}({\mathcal{P}})$. On the other hand, RE of any order are Schur-concave
functions~\cite{Zyczkowski}. In fact, every function $f({\mathcal{P}})$ which is Schur concave
can represent a reasonable measure of information,
since it is maximized by a uniform probability distribution, while
minimum is provided with concentrated distributions ${\mathcal{P}} = \{p_i =1, p_{j\neq i} = 0\}$.
Some further properties can be found, e.g., in Refs.~\cite{Re2,Re1,PJ1}.

Note particularly that RE of two {\em independent} experiments (sources)
is additive. In fact, it was proved in Ref.~\cite{Re2}
that RE is the most general information measure compatible with additivity of
independent information and Kolmogorov axioms of probability
theory.
%
%%%%%%%%%%%%%%%%%%%%%%%%%%%%%%%%%%%%%%%%%%%%%%%%%%%%%%%%%%%%%%%%%
\subsection{THC entropy: entropy of long distance correlated
systems}\label{V}
%%%%%%%%%%%%%%%%%%%%%%%%%%%%%%%%%%%%%%%%%%%%%%%%%%%%%%%%%%%%%%%%%

THC entropy was originally introduced in 1967 by Havrda and
Charv\'{a}t in the context of information theory of computerized
systems~\cite{HaCh} and together with the $\alpha$-norm entropy
measure~\cite{Arimito1} it belongs to class of pseudo-additive
entropies. In contrast with R\'{e}nyi's or Shannon's entropy THC
entropy does not have (as yet) an operational characterization.
Havrda--Charv\'{a}t structural entropy, though quite well known
among information theorists, had remained largely unknown in
physics community. It took more than two decades till Tsallis in
his pioneering work~\cite{Ts1} on generalized (or non-extensive)
statistics rediscovered this entropy. Since then THC entropy
has been employed in many physical systems. In this connection one
may particularly mention, Hamiltonian systems with long-range
interactions, granular systems, complex networks, stock market
returns, etc.. For recent review see, e.g., Ref.~\cite{Tsallis_book}.

In the case of a discrete distribution ${\mathcal{P}} = \{ p_1,
\ldots, p_n \}$ the THC entropy takes the form:

\begin{eqnarray}
{\mathcal{S}}_q({\mathcal{P}}) = \frac{1}{(1-q)} \left[ \sum_{k=1}^n
(p_k)^q  -1\right]\, , \;\;\;\; q> 0\, .
\end{eqnarray}
\noindent Various axiomatic treatments of THC entropy were proposed in the literature.
 For our purpose the most
convenient set of axioms is the following~\cite{Ab2}:
\vspace{3mm}
\begin{enumerate}
\item For a given integer $n$ and given ${\mathcal{P}} = \{ p_1,
p_2, \ldots , p_n\}$ ($p_k \geq 0, \sum_k^n p_k =1$),
${\mathcal{S}}({\mathcal{P}})$ is a continuous with respect to all
its arguments.

\item For a given integer $n$, ${\mathcal{S}}(p_1, p_2, \ldots ,
p_n)$ takes its largest value for $p_k = 1/n$ ($k=1,2, \ldots,
n$).
%${\mathcal{I}}_{q}({\mathcal{P}})$ takes its largest value for
%$p_k = 1/n, (k = 1,2, \ldots, n)$, i.e., the gained information is
%largest when we known least about the original system.

\item For a given $q\in {\mathbb{R}}$; ${\mathcal{S}}(A\cup B) =
{\mathcal{S}}(A) +
{\mathcal{S}}(B|A) + (1-q){\mathcal{S}}(A){\mathcal{S}}(B|A)$
with\\[2mm]
 $\mbox{\hspace{2.5cm}}{\mathcal{S}}(B|A) =
\sum_k \varrho_k(q) \ {\mathcal{S}}(B|A=A_k)$,\\[2mm]
and $\varrho_k(q) = p_k^q/\sum_k p_k^q$\, (distribution
${\mathcal{P}}$ corresponds to the experiment $A$).

%For independent events, i.e., ${\mathcal{R}} = {\mathcal{P}}\times
%{\mathcal{Q}}=
%\{p_iq_k \}$: \\
%${\mathcal{I}}_{q}({\mathcal{P}}\times {\mathcal{Q}}) =
%{\mathcal{I}}_{q}({\mathcal{P}}) + {\mathcal{I}}_{q}({\mathcal{Q}})$.
%
%\item In general case there is a continuous invertible function
%$f(x)$ such that: ${\mathcal{I}}_{q}\partial
%f({\mathcal{I}}_{q})/\partial
%{\mathcal{I}}_{q} = g({\mathcal{I}}_{q}) f({\mathcal{I}}_{q})$.\\
%\\
%Here ${\mathcal{I}}_{q}({\mathcal{P}}\cup {\mathcal{Q}}) =
%{\mathcal{I}}_{q}({\mathcal{I}}_{q}({\mathcal{P}}),
%{\mathcal{I}}_{q}(\mathcal{Q}))$. The scaling function $g(x)$ is
%common to both ${\mathcal{I}}_{q}({\mathcal{P}}\cup {\mathcal{Q}}),
%{\mathcal{I}}_{q}({\mathcal{P}})$ and ${\mathcal{I}}_{q}({\mathcal{Q}})$
%.

\item ${\mathcal{S}}(p_1,p_2, \ldots , p_n, 0 ) =
{\mathcal{S}}(p_1,p_2, \ldots , p_n)$.
\end{enumerate}
\vspace{3mm}
As we said before, one keeps here the linear mean but generalizes the
additivity law. In fact, the additivity law in axiom~3 is nothing
but the Jackson sum
%(or $q$--additivity)
known from the $q$-calculus~\cite{FJ1}; there one defines the
Jackson basic number $[X]_{\{q\}}$ of quantity $X$ as
\begin{eqnarray}
[X]_{\{q\} } = (q^X -1)/(q-1) \ \Rightarrow  \  [X+Y]_{\{q\}} =
[X]_{\{q\}} + [Y]_{\{q\}} + (q-1)[X]_{\{q\}}[Y]_{\{q\}}\, .
\end{eqnarray}
The connection with axiom~3 is then established when $q
\rightarrow (2-q)$. Nice feature of the $q$-calculus is that it
formalizes many mathematical manipulations. For instance, using
the $q$-logarithm
\begin{eqnarray}
\ln_{\{q\}} x \ = \ - \ln_{\{2-q\}}\left(\frac{1}{x} \right) \ = \
\frac{1}{1-q} \ (x^{1-q} -1) \, , \label{IIC6}
\end{eqnarray}
\noindent THC entropy can be concisely written as the $q$-deformed
Shannon's entropy, i.e.,
\begin{eqnarray}
{\mathcal{S}}_q({\mathcal{P}}) \ = \ - \sum_{k=1}^n p_k
\ln_{\{2-q\}} p_k \ = \ - \sum_{k=1}^n p_k^q \ln_{\{q\}} p_k   \ =
\ \sum_{k=1}^n p_k \ln_{\{q\}} \left( \frac{1}{p_k} \right)\, .
\end{eqnarray}
Some elementary properties of ${\mathcal{S}}_q$ are positivity, concavity (and
Schur concavity) for all values of $q$ and indeed non-extensivity. There hold
also inequalities between all three entropies, namely:
\begin{equation}{\mathcal{H}}({\mathcal{P}}) \ \leq \
{\mathcal{I}}_q({\mathcal{P}}) \ \leq \
{\mathcal{S}}_q({\mathcal{P}})\, ,
\end{equation}
for $0 < q \leq  1$, and
\begin{equation}{\mathcal{S}}_q({\mathcal{P}}) \ \leq \
{\mathcal{I}}_q({\mathcal{P}}) \ \leq
\ {\mathcal{H}}({\mathcal{P}})\, ,
\end{equation}
for $q \ \geq \ 1$. For a monograph that cover this subject in more depth the reader is
referred to Ref.~\cite{Ts2a}.

%%%%%%%%%%%%%%%%%%%%%%%%%%%%%%%%%%%%%%%%%%%%%%%%%%%
\section{J-A axioms and solutions}\label{VII}
%%%%%%%%%%%%%%%%%%%%%%%%%%%%%%%%%%%%%%%%%%%%%%%%%%%

It would be conceptually desirable to have a unifying axiomatic
framework in which both properties of R\'{e}nyi and THC entropies are both represented.
In Ref.~\cite{PJ3} one of us proposed the following {\em
natural} synthesis of the previous two axiomatics:
\begin{enumerate}
\item For a given integer $n$ and given ${\mathcal{P}} = \{ p_1,
p_2, \ldots , p_n\}$ ($p_k \geq 0, \sum_k^n p_k =1$),
${\mathcal{D}}({\mathcal{P}})$ is a continuous with respect to all
its arguments.

\item For a given integer $n$, ${\mathcal{D}}(p_1, p_2, \ldots ,
p_n)$ takes its largest value for $p_k = 1/n$ ($k=1,2, \ldots,
n$).
%${\mathcal{I}}_{q}({\mathcal{P}})$ takes its largest value for
%$p_k = 1/n, (k = 1,2, \ldots, n)$, i.e., the gained information is
%largest when we known least about the original system.

\item For a given $q\in {\mathbb{R}}$; ${\mathcal{D}}(A\cup B) =
{\mathcal{D}}(A) +
{\mathcal{D}}(B|A) + (1-q){\mathcal{D}}(A){\mathcal{D}}(B|A)$
with\\[2mm]
 $\mbox{\hspace{0.5cm}}{\mathcal{D}}(B|A) =
f^{-1}\left(\sum_k \varrho_k(q) \
f\left({\mathcal{D}}(B|A=A_k)\right)\right)$,\\[2mm]
and $\varrho_k(q) = p_k^q/\sum_k p_k^q$\, (distribution
${\mathcal{P}}$ corresponds to the experiment $A$). Function $f$ is invertible and positive in $[0, \infty)$.

%For independent events, i.e., ${\mathcal{R}} = {\mathcal{P}}\times
%{\mathcal{Q}}=
%\{p_iq_k \}$: \\
%${\mathcal{I}}_{q}({\mathcal{P}}\times {\mathcal{Q}}) =
%{\mathcal{I}}_{q}({\mathcal{P}}) + {\mathcal{I}}_{q}({\mathcal{Q}})$.
%
%\item In general case there is a continuous invertible function
%$f(x)$ such that: ${\mathcal{I}}_{q}\partial
%f({\mathcal{I}}_{q})/\partial
%{\mathcal{I}}_{q} = g({\mathcal{I}}_{q}) f({\mathcal{I}}_{q})$.\\
%\\
%Here ${\mathcal{I}}_{q}({\mathcal{P}}\cup {\mathcal{Q}}) =
%{\mathcal{I}}_{q}({\mathcal{I}}_{q}({\mathcal{P}}),
%{\mathcal{I}}_{q}(\mathcal{Q}))$. The scaling function $g(x)$ is
%common to both ${\mathcal{I}}_{q}({\mathcal{P}}\cup {\mathcal{Q}}),
%{\mathcal{I}}_{q}({\mathcal{P}})$ and ${\mathcal{I}}_{q}({\mathcal{Q}})$
%.

\item ${\mathcal{D}}(p_1,p_2, \ldots , p_n, 0 ) =
{\mathcal{D}}(p_1,p_2, \ldots , p_n)$.
\end{enumerate}
Note particularly that due to the non-linear nature of the
non-additivity condition there is no need to select a
normalization condition for ${\mathcal{D}}_q$. In Ref.~\cite{PJ3} it was shown that above axioms allow for only one class of solutions, leading to
an entirely new family of physically conceivable entropy functions. For reader's convenience are the basic steps of the proof sketched in~\ref{AppA}. In particular, the resulting hybrid entropy has the following form:
%
%\begin{eqnarray}
%{\mathcal{D}}_q^{\beta}({\mathcal{P}}) =
%\frac{1}{(1-q)(1-\beta)}\left[ 1 - \left( \sum_k (p_k)^q
%\right)^{1-\beta} \right]\, .
%\end{eqnarray}
%
%\noindent Surprisingly enough, for $\beta \in [0.5, 0.77]$ this
%coincides with Alexander's (active) information used in
%neurophysics and biophysics (DNA sampling)~\cite{}.

\begin{eqnarray}
{\mathcal{D}}_q(A) \ = \  \frac{1}{1-q}\ \left( e^{-(1-q)^2
d{\mathcal{I}}_q/dq} \sum_{k=1}^n (p_k)^q -1 \right)\ = \ \ln_{\{q\}}  e^{-  \langle \ln {\mathcal{P}} \rangle_q} \, .
\label{IIIh}
\end{eqnarray}

Let us further remark that axiom 4 restricts the possible values
of $q$ to $q \geq 0$. This is because  ${\mathcal{D}}_q$ would
otherwise tend to infinity if some of $p_k$ would tend to zero.
The latter would be counter-intuitive, because without changing the
probability distribution we would gain an infinite information.
Value $q=0$ must be also ruled out on the basis of axiom 2,
because ${\mathcal{D}}_0$ would yield an expression not dependent
on the probability distribution ${\mathcal{P}}$ but only on the
number of outcomes (or events) --- i.e., ${\mathcal{D}}_0$ would
be a system (source) insensitive. In addition, by further analysis in \ref{AppA}, supported by the concept of Schur-concavity in Section~\ref{sec: concavity}
we show that
${\mathcal{D}}_q$ is well-defined only for $q \geq \frac{1}{2}$. In particular, for $q < \frac{1}{2}$ the entropy ${\mathcal{D}}_q$ has a local minimum at ${\mathcal{P}} = \{1/n,\dots,1/n\}$ (rather than maximum) and therefore it does not fulfill axiom 2.
% More generally, maximality axiom is used in
% derivation of hybrid entropy only for special values and one has to additionally
% show that the entropy does not violate axiom 2. The usual procedure is to show that
% the entropy is a concave function of probability distributions. Nevertheless, this is only sufficient condition
% and as known from example of R\'{e}nyi entropy, there are entropies which are not concave and still well defined.
% In this connection, there exist an alternative property that guarantees validity of maximality axiom, namely Schur-concavity.
% Concavity properties of hybrid entropy are discussed in Section~\ref{sec: concavity}. Back to maximality axiom, in \ref{AppA} is shown that
% hybrid entropy is well-defined for $q \geq \frac{1}{2}$. For $q < \frac{1}{2}$ has ${\mathcal{D}}_q$ local minimum in $u = (1/n,\dots,1/n)$
% and therefore does not fulfill axiom 2.
Some basic properties of the hybrid entropy ${\mathcal{D}}_q$ are presented in~\ref{AppB}.

\begin{figure}[t]
\begin{center}
\includegraphics[width=5.3cm]{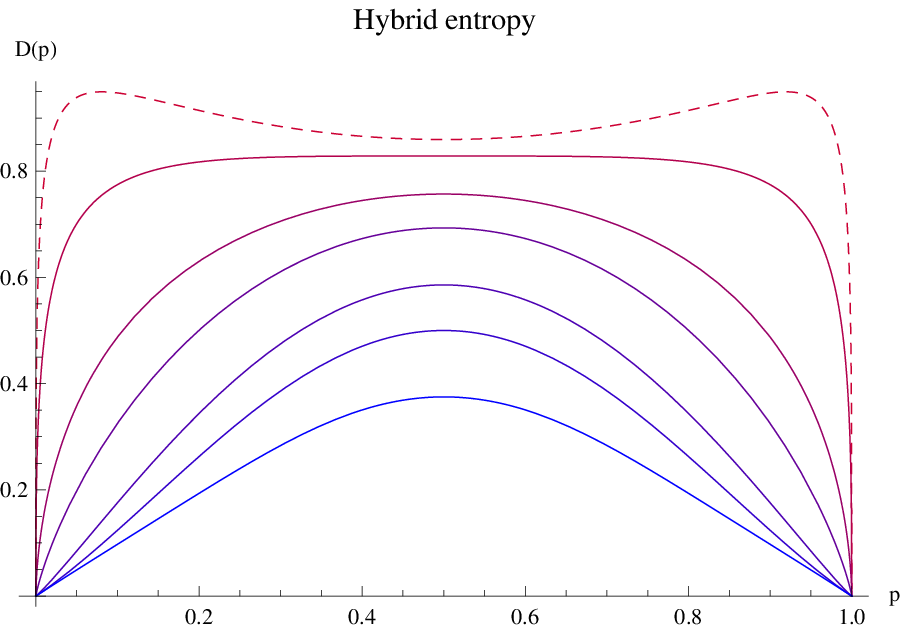}
\includegraphics[width=5.3cm]{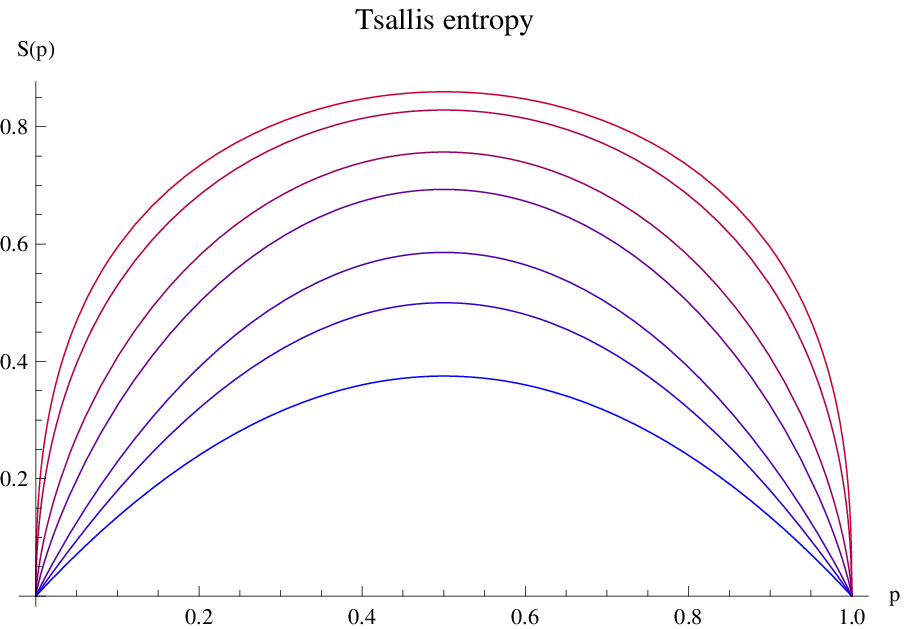}
\includegraphics[width=5.3cm]{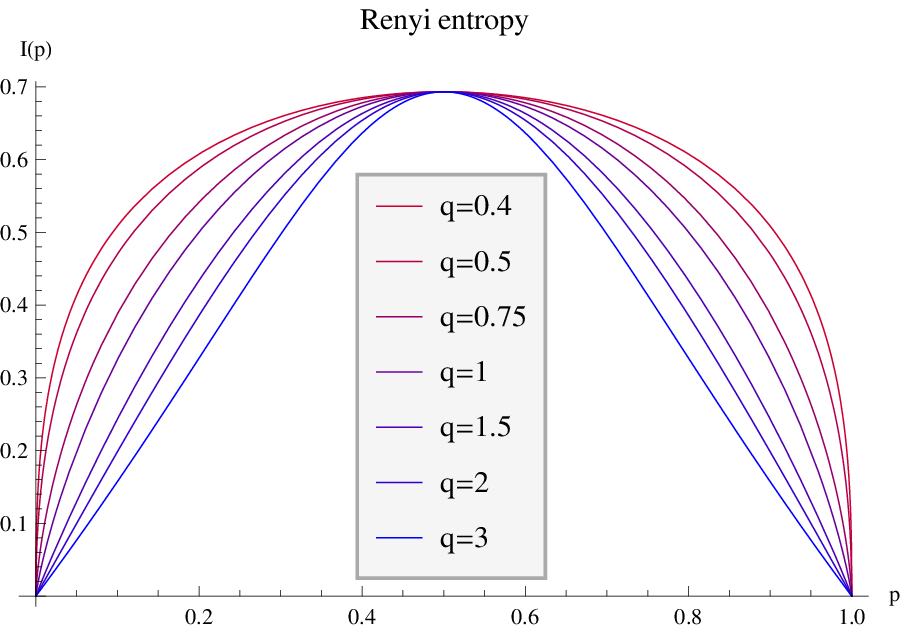}
\end{center}
\caption{Comparison of entropies for several values of $q$ for two-event systems ($\mathcal{P} = \{p,1-p\}$). The dashed curve represents the hybrid entropy ${\mathcal{D}}_{_{0.4}}$ which violates the maximality axiom.}
\label{Fif.3bb}
\end{figure}

Before studying further implications of the formula (\ref{IIIh}),
there are two immediate consequences which warrant special
mention. The first is that, from the condition
$d{\mathcal{I}}_q/dq \leq 0$ (see Section \ref{III}) we
have
\begin{eqnarray}
{\mathcal{D}}_q(A) = \left\{
\begin{array}{ll}   \geq  \ {\mathcal{S}}_q(A) \,
 & \mbox{if $q \leq 1$}\\
\leq \ {\mathcal{S}}_q(A) \, & \mbox{if $q \geq 1$}
\end{array} \right. \, ,
\end{eqnarray}
where equality holds, if and only if,  $q=1$ or
$d{\mathcal{I}}_q/dq =0$. These mean that either
${\mathcal{D}}_q(A)$ and ${\mathcal{S}}_q(A)$ jointly coincide
with Shannon's entropy or that ${\mathcal{P}}$ is uniform or $\{
1,0, \ldots, 0\}$. Hence, combining this with inequalities between THC, R\'{e}nyi end Shannon entropy, we obtain
\begin{eqnarray}
&&0 \ \leq \ {\mathcal{H}}({\mathcal{P}}) \ \leq \
{\mathcal{I}}_q({\mathcal{P}}) \ \leq \
{\mathcal{S}}_q({\mathcal{P}})\ \leq \
{\mathcal{D}}_q({\mathcal{P}}) \ \leq \ \ln_{\{q\}} n \,\,\,\,\,\,\,\,\,
\mbox{for} \,\, \frac{1}{2} \leq q \leq 1\, ,\nonumber \\  &&  0 \ \leq \
{\mathcal{D}}_q({\mathcal{P}}) \ \leq \
{\mathcal{S}}_q({\mathcal{P}}) \ \leq \
{\mathcal{I}}_q({\mathcal{P}})\ \leq \
{\mathcal{H}}({\mathcal{P}}) \ \leq \ \ln n
\,\,\,\,\,\,\,\,\,\,\,\, \mbox{for} \,\, q \ \geq \ 1\, .
\label{IIIl}
\end{eqnarray}
The result (\ref{IIIl}) implies that by investigating the
information measure ${\mathcal{D}}_q$ with $q < 1$ we receive more
information than restricting our investigation just to entropies
${\mathcal{I}}_q$ or ${\mathcal{S}}_q$. On the other hand, when $q
> 1$ then both ${\mathcal{I}}_q$ and ${\mathcal{S}}_q $ are more
informative than ${\mathcal{D}}_q$. The first set of inequalities is also valid for $q<\frac{1}{2}$, but the
last relation to $\ln_{\{q\}}n$ is not true for the hybrid entropy. The practical illustration of the above inequalities can be seen in Fig.~\ref{Fif.3bb} for simple distribution ${\mathcal{P}} = \{p,1-p\}$.

In practical cases one usually requires more than one $q$ to gain more complete information about the system. In fact, when entropies ${\mathcal{I}}_q$ or
${\mathcal{S}}_q$ are used, it is necessary to know them for all
$q$ in order to obtain a full information on a given statistical
system~\cite{PJ1}. For ensuing applications in strange attractors the
reader may consult Ref.~\cite{HaPr1}, for reconstruction theorems
see, e.g., Refs.~\cite{Re2,PJ1}.

The second comment to be made concerns the fact that when the
statistical system in question is a multifractal\footnote{The
necessary essentials on multifractals are presented in~\ref{AppC}.} then relations (\ref{B2})-(\ref{B4}) assert that
\begin{eqnarray}
(1-q)^2 \frac{d {\mathcal{I}}_q}{dq}\ = \ (a- f(a))
\ln\varepsilon  \ = \ \ln \left(\sum_k^{N(a)} \!
p_k(\varepsilon)\right) \, , \label{IIIn}
\end{eqnarray}
where summation runs only over support boxes of size $\varepsilon$
with the scaling exponent $a$. Alternatively, we could have
started with the first relation in  Eq.~(\ref{IIIg}) and use the
multifractal canonical relations (see Ref.~\cite{PJ1}) in which
case the result would have been again (\ref{IIIn}). So for the
coarse-grained multifractal with the mesh size $\varepsilon$ the
corresponding entropy ${\mathcal{D}}_q$ reads
\begin{eqnarray}
{\mathcal{D}}_q(A) \ = \ \frac{1}{(1-q)} \left(\frac{
\sum_{k=1}^n(p_k(\varepsilon))^q}{\sum_k^{N(a)} \!\!
p_k(\varepsilon)} -1 \right)
%\ = \ \frac{1}{(1-q)} \left(
%\frac{\sum_{k=1}^n(p_k(\varepsilon))^q}{N(a)P(a)} - 1\right)
\, .
\label{3.13.b}
\end{eqnarray}
Now, the passage from multifractals to single--dimensional
statistical systems is done by assuming that the $a$-interval is
infinitesimally narrow and that PDF is smooth~\cite{PJ1,Cv}. In
such a case Cvitanovic's condition~\cite{Cv} holds, namely  both
$a$ and $f(a)$ collapse to $a = f(a) \equiv D$ and $q = f'(a) =1$.
So, for example, for a statistical system with a smooth PDF and
the support space ${\mathbb{R}}^d$ the relation (\ref{3.13.b}) implies that the entropy ${\mathcal{D}}_q$
coincides with Shannon's ${\mathcal{H}}$. In this connection it is important to stress that
the similarity of (\ref{3.13.b}) with THC entropy is only apparent. In order to have THC entropy one needs to have $N(a) = n$, i.e.,
the entire probability measure must be accumulated around the unifractal with the scaling exponent $a$.  According to the Billingsley (or curdling) theorem~\cite{Man1,Bill} this is possible only when $a = f(a) = D$, i.e., only when ${\mathcal{D}}_q = {\mathcal{H}}$.
As a byproduct of
Eq.~(\ref{IIIl}) we may notice that for single-dimensional systems
with smooth PDF's ${\mathcal{S}}_q$ and ${\mathcal{I}}_q$ must
approach Shannon's entropy~\cite{PJ1}. We remark that this may
help to understand why Shannon's entropy plays such a predominant
r\^{o}le in physics of single-dimensional sets.

In what follows, we examine the class of distributions that
represent maximizers for ${\mathcal{D}}_q(A)$ subject to
constraint imposed by the average value of energy.

%${\mathcal{D}}_q$ is a monotonic function of the escort entropy
%
%\begin{eqnarray}
%{\mathcal{S}}_q^e({\mathcal{P}}) = \sum_k \rho_k(q) \ln p_k\, ,
%\end{eqnarray}
%
%but contrast to it ${\mathcal{D}}_q$ has better concavity
%properties for ${\mathcal{P}}$. Indeed, it is simply to check that
%${\mathcal{D}}_q$ is concave when
%
%\begin{eqnarray}
%
%\end{eqnarray}
%
%
%
%when we restrict values of ${\mathcal{P}}$ by the constraint
%
%\begin{eqnarray}
%\frac{d{\mathcal{I}}_q({\mathcal{P}})}{dq} \ = \ \max_{p_i}\frac{
%{\mathcal{I}}_q({\mathcal{P}})}{(1-q)} \ = \ \frac{\ln n}{(1-q)}\, .
%\label{IIIk}
%\end{eqnarray}
%
%(Because $d{\mathcal{I}}_q/dq \leq 0$,  Eq.(\ref{IIIk}) can be
%fulfilled only for $q>1$.) If the ****
%
%Under assumption (\ref{IIIk}) Eq.(\ref{IIIh}) reduces to
%
%\begin{eqnarray}
%{\mathcal{D}}_q(A) =  \frac{1}{1-q}\ \left( n^{q-1} \sum_{k=1}^n
%(p_k)^q -1 \right) \equiv - {\mathcal{C}}_q(A)\, . \label{IIIi}
%\end{eqnarray}
%
%The reader may recognize in ${\mathcal{C}}_q$ the generalized
%measure of cross--entropy\footnote{Known also as Csisz\'{a}r's
%measure of directed divergence~\cite{Csi2}.} of Havrda and
%Charvat~\cite{HaCh,Kapur} used in mathematical statistics and
%communication theory. For $q=2$ we recover Neyman's $\chi^2$
%measure. In passing we note that ${\mathcal{D}}_q \geq 0$ for
%$\forall q \in \mathbb{R}$ and
%$\lim_{q\rightarrow1}{\mathcal{D}}_q  = {\mathcal{I}}_1 =
%{\mathcal{S}}_1$.

%%%%%%%%%%%%%%%%%%%%%%%%%%%%%%%%%%%%%%%%%%%%
\section{MaxEnt distribution}\label{X}
%%%%%%%%%%%%%%%%%%%%%%%%%%%%%%%%%%%%%%%%%%%%

According to information theory, the MaxEnt principle
yields distributions which reflect least bias and maximum
uncertainty about information not provided to a recipient (i.e.,
observer). Important feature of the usual Gibbsian MaxEnt
formalism is that the maximal value of entropy is a concave
function of the values of the prescribed constraints (moments),
and maximizing probabilities are all grater than zero~\cite{HBC}.
The first is important for thermodynamical stability and the
second for mathematical consistency. In this section we will see
that both mentioned features hold true also in the case of the
${\mathcal{D}}_q$ entropy.

Let us first address the issue of maximizers for
${\mathcal{D}}_q$. To this end we
%will consider two
%different ways of implementing the constraints. We
shall seek the conditional extremum of ${\mathcal{D}}_q$ subject
to the constraints imposed by the averaged value of energy $E$ (or
generally any random quantity representing the constant of the
motion) in the form
\begin{eqnarray}
\langle E \rangle_r \ = \ \sum_k \varrho_k(r)E_k\, .
\end{eqnarray}
For the future convenience we initially keep $r$ not necessary
coincident with $q$. Taking into account the normalization
condition for $p_i$ we ought to  extremize the functional
\begin{eqnarray}
L_{q,r}({\mathcal{P}}) \ = \ {\mathcal{D}}_q({\mathcal{P}}) - \Omega
\ \frac{\sum_k (p_k)^r E_k}{\sum_k (p_k)^r} - \Phi \sum_k p_k\, ,
\label{Va1}
\end{eqnarray}
with $\Omega$ and $\Phi$ being the Lagrange multipliers. Setting
the derivatives of $L_{q,r}({\mathcal{P}})$ with respect to $p_1,
p_2 \ldots,$ etc., to zero, we obtain
\begin{eqnarray}
\frac{\partial L_{q,r}({\mathcal{P}})}{\partial p_i} \ &=& \
e^{(q-1)\sum_k \varrho_k(q) \ln p_k} \left[q
(\langle \ln {\mathcal{P}}\rangle_q -\ln p_i ) -1 \right]
\frac{(p_i)^{q-1}}{\sum_k (p_k)^q} \nonumber \\
&-& r\Omega \left( E_i - \langle E \rangle_r \right)
\frac{(p_i)^{r-1}}{\sum_{k} (p_k)^r} - \Phi \ = \ 0 \, ,
\;\;\;\;\; i = 1,2, \ldots , n\, . \label{VIIIa}
\end{eqnarray}
Note that when both $q$ and $r$ approach $1$, (\ref{VIIIa})
reduces to the usual condition for Shannon's maximizer. This, in
turn, ensures that in the $(q,r)\rightarrow(1,1)$ limit the
maximizer of (\ref{Va1}) is Gibbs's canonical-ensemble distribution.
Let us now concentrate on the two most relevant situations, namely when $r=q$ and $r=1$.

%%%%%%%%%%%%%%%%%%%%%%%%%%%%%%%%%%%%%%%%%%%%%%%%%%%%%%%%%%%%%%
\subsection{ The $r=q$ case}
%%%%%%%%%%%%%%%%%%%%%%%%%%%%%%%%%%%%%%%%%%%%%%%%%%%%%%%%%%%%%%

\vspace{2mm}

When we decide to use $r=q$ (i.e., when the non-linear moment
constraints are implemented via escort distribution) it follows
from (\ref{VIIIa}) that
\begin{eqnarray}
\Phi  (p_i)^{1-q} \sum_k (p_k)^q \ = \
e^{(q-1)\langle \ln {\mathcal{P}} \rangle_q }
\left[q\left(\langle \ln {\mathcal{P}} \rangle_q - \ln p_i\right)
-1\right] - q \Omega (E_i - \langle E\rangle_q ) \, .
\label{VIIIb}
\end{eqnarray}
Multiplying both sides of (\ref{VIIIb}) by $\varrho_i(q)$, summing
over $i$ and taking the normalization condition $\sum_k p_k =1$ we
obtain
\begin{eqnarray}
\Phi = -  e^{(q-1)\langle \ln {\mathcal{P}} \rangle_q } \; \;
\Rightarrow \; \; \frac{\ln(-\Phi)}{q-1} \ = \ \langle \ln
{\mathcal{P}} \rangle_q \, \;\; \Rightarrow \; \;
{\mathcal{D}}_q({\mathcal{P}})|_{\mbox{\footnotesize{max}}} \ = \
\frac{1}{q-1}\ ( \Phi + 1) \, . \label{VIIIc}
\end{eqnarray}
%
%So wee see that the normalization of probability puts rather strong
%restriction on possible distributions that can be used. Note that
%analogous condition occurs also in the case when Shannon's entropy
%is extremized.
Plugging result (\ref{VIIIc}) back into (\ref{VIIIb})
we obtain after some algebra
\begin{eqnarray}
\sum_k (p_k)^q \  =  \ (p_i)^{q-1} \left[ q \ln p_i + \left( 1 -
\frac{q \ln(-\Phi)}{q-1} - \frac{q\Omega}{\Phi} \ (E_i - \langle E
\rangle_q) \right) \right] \, , \label{V1}
\end{eqnarray}
which must be true for any index $i$. On the substitution
\begin{eqnarray}
{\mathcal{E}}_i \ =  \ 1 - \frac{q \ln(-\Phi)}{q-1} -
\frac{q\Omega}{\Phi} \ (E_i - \langle E \rangle_q)\, ,
\end{eqnarray}
this leads to the equation
\begin{eqnarray}
\kappa (p_i)^{1-q} \ = \ q \ln p_i \ + \ {\mathcal{E}}_i\, .
\label{V4}
\end{eqnarray}
Here we have denoted $\sum_k (p_k)^q \equiv \kappa$. Equation
(\ref{V4}) has the solution
\begin{eqnarray}
p_i \ = \ \left[  \frac{q}{\kappa (q-1)} \ W\left( \frac{\kappa
(q-1)}{q} \ e^{(q-1) {\mathcal{E}}_i/q}\right) \right]^{1/(1-q)}
 \ =
\ \exp\left\{ \frac{W\left( \frac{\kappa (q-1)}{q} \ e^{(q-1)
{\mathcal{E}}_i/q} \right)}{(q-1)} - {\mathcal{E}}_i/q \right\}\,
, \label{V2}
\end{eqnarray}
with $W(x)$ being the Lambert--W function~\cite{CGHJK}.

A couple of comments are now in order. First, $p_i$'s as
prescribed by (\ref{V2}) are positive for any value of $q > 0$.
This is a straightforward consequence of the following two
identities~\cite{CGHJK}:
\begin{eqnarray}
&&W(x) \ = \ \sum_{n=1}^{\infty} \frac{(-1)^{n-1}
n^{n-2}}{(n-1)!}\
x^n \, , \label{V3a} \\
&& W(x) \ = \ x \ e^{-W(x)}\, . \label{V3b}
\end{eqnarray}
Indeed, Eq.~(\ref{V3a}) ensures that for $x < 0$ also $W(x) < 0$
and hence $W(x)/x > 0$. Thus for $0 < q <1$ the positivity of
$p_i$'s is a simple consequence of the first part of (\ref{V2}).  Positivity for $q \geq 1$ follows directly
from the relation (\ref{V3b}) and the second part of (\ref{V2}).

Second, as $q  \rightarrow 1$ the
entropy ${\mathcal{D}}_q \rightarrow {\mathcal{H}}$ and we expect
that $p_i$'s defined by (\ref{V2}) should approach the Gibbs
canonical-ensemble distribution in this limit. To see that
this is indeed the case, let us note that
\begin{eqnarray}
\left.\Phi\right|_{q =  1} \ = \  -1\, , \;\;\;
\left.{\mathcal{E}}_i\right|_{q = 1 } \ = \ 1 + {\mathcal{H}} +
\Omega (E_i - \langle E \rangle )\, , \;\;\;\mbox{and} \;\;\;
\left.\kappa \right|_{q = 1}  \ = \ 1\, .
\end{eqnarray}
Then
\begin{eqnarray}
\left.p_i\right|_{q = 1} \ = \ e^{1 - [1 + {\mathcal{H}} + \Omega
(E_i - \langle E \rangle)] } \ = \ e^{\Omega F - \Omega E_i}  \ =
\ e^{-\Omega E_i}/Z \, , \label{IVA44}
\end{eqnarray}
(here $F$ is the Helmholtz free energy) which after identification $\left.\Omega\right|_{q = 1} = \beta$
leads to the desired result. Note also that (\ref{V2}) is invariant
under uniform translation of the energy spectrum, i.e., the
corresponding $p_i$ is independent of the choice of the energy
origin.

Third, there are situations, when Eq.~\eqref{V4} has no solution, or it gives solution for $p_i \not \in [0,1]$.
To see this, we may notice that when $q>1$, the left-hand side of \eqref{V4} is greater than $\kappa$,
from which follows that ${\mathcal{E}}_i \geq \kappa$ for all $i$'s.
For $q<1$ the left-hand side of \eqref{V4} acquires values from $[0,\kappa]$ which (after using the fact that $q< 1 < \kappa$) leads again to the condition
${\mathcal{E}}_i \geq \kappa$. In both cases are therefore ${\mathcal{E}}_i$ positive. Thus, for energies, for which
$\Delta_q E_i = E_i - \langle E\rangle_q$ is too negative,
Eq.~\eqref{V4} has no solution, and the corresponding occupation probability is zero.
Contrary to MaxEnt distributions of other commonly used  entropies, there exist energy levels here,
for which MaxEnt distributions of ${\mathcal{D}}_q$ have zero occupation probabilities.
This might provide a natural conceptual playground for statistical systems with energy gaps (e.g., disordered systems,
carbon nanotubes)
or for system with various super-selection rules (e.g., first-quantized relativistic systems).

Finally, there does not seem to by any simple method for a unique
determination of $\Phi$ and $\Omega$ from the constraint
conditions\footnote{In conventional statistical physics one does not solve $\Omega$ in terms of averaged energy (i.e., internal energy $U$) since $\Omega$ can be identified with inverse temperature which is much more fundamental quantity than $U$. In fact, it is $U$ that is typically given as a function of $\Omega$. }. In fact, only asymptotic situations for large and vanishingly small $\Omega$ can be successfully resolved (this will be relegated to Sections~\ref{4.1.2a} and \ref{4.1.3a}). There exists, however, systems of a practical interest --- namely
multifractal systems, where we can
give to relations (\ref{V2}) a very satisfactory physical
interpretation, without resolving $\langle E \rangle_q$ in terms
of $\Phi$ and $\Omega$.

%%%%%%%%%%%%%%%%%%%%%%%%%%%%%%%%%%%%%%%%%%%%%%%%%%%%%%%%%%%%%%
\subsubsection{Multifractal case}
%%%%%%%%%%%%%%%%%%%%%%%%%%%%%%%%%%%%%%%%%%%%%%%%%%%%%%%%%%%%%%

\vspace{2mm}
In case when  a statistical system under investigation fits the multifractal paradigm\footnote{For a brief introduction to multifractals see~\ref{AppC}.}, we can cast Eq.~\eqref{V4} in the form
\begin{equation}\label{M1}
\varepsilon^{\tau(q) + a_i (1-q)} \ \sim  \ 1 \ + \ q \left[a_i - \langle a
\rangle_q(\varepsilon) \right]\left(1 + \frac{\Omega}{\Phi} \right) \ln
\varepsilon \, ,
\end{equation}
where $\tau$ and $a_i$ are correlation exponent and Lipshitz--H{\"{o}}lder exponent, respectively. Note that the $q$-mean $\langle a \rangle_{q}(\varepsilon)$ at the coarse-grained
scale $\varepsilon$ is proportional to the $q$-mean of log-PDF, namely
\begin{equation}
\langle \ln {\mathcal{P}} \rangle_q \ = \ {\sum_k \varrho_k(q) \ln p_k} \ \sim \ {\sum_k \varrho_k(q) a_k \ln \varepsilon} \ = \ \langle a \rangle_q(\varepsilon) \ln \varepsilon \, .
\end{equation}
So, in particular,  $\Phi = - \varepsilon^{(q-1) \langle a \rangle_q}$ as can be directly deduced from Eq.~(\ref{VIIIc}).

Equation~\eqref{M1} has several important implications.
Firstly, we remind  the reader that in the long-wave limit
(i.e., when $\varepsilon \rightarrow 0$), one can use analogy with ordinary statistical thermodynamics and interpret $\langle a \rangle_{q}$ as the most likely value of ``energy'' of a system
immersed in a heat bath with the effective inverse temperature $\beta = q$ (see, e.g., Ref.~\cite{PJ1}).
This is a version of the {Billingsley} (or {curdling}) theorem~\cite{Man1,Bill,PJ2}, which states that the Hausdorff dimension of the set on which the escort probability $\varrho_k(q)$ is concentrated is $f(\langle a \rangle_{q}) = q \langle a \rangle_{q} - \tau(q)$.
In addition, the relative probability of the complement set approaches zero when $\varepsilon \rightarrow 0$.
%i.e., thermodynamic limit,
This in turn means that for each $q$ there exists one scaling exponent, namely $a_i = \langle a \rangle_{q}$ which dominates, e.g., the partition function $\kappa$, whereas $p_i$'s with other Lipshitz--H{\"{o}}lder exponents have only marginal contribution.
% In this case we have that
%%
%\begin{equation}
%a_i \ = \ \frac{\tau(q)}{q-1} \ = \ D_q\, ,
%\end{equation}
%%
%which is property inherited from R\'{e}nyi's entropy.
%Indeed, in thermodynamic limit is possible to observe only a part of the whole multifractal system the scales with the scaling exponent $\langle a \rangle_q$ and the choice of $q$ influences on which part are we focused.

Note that the aforesaid indeed mimics the situation occurring in equilibrium statistical physics.
There, in the canonical formalism one works with (usually infinite)
ensemble of identical systems with all possible energy configurations. But only
the configurations with $E_i \sim \langle E \rangle_{\beta}$
dominate partition function in the thermodynamic limit. Choice of temperature $T = 1/\beta$
then prescribes the  contributing energy configurations.

Secondly, for small $\varepsilon$ we have
\begin{eqnarray}
\tau(q) + a_i (1-q) \ \sim \ \ln\left\{ 1 \ + \ q \left[a_i - \langle a
\rangle_q(\varepsilon) \right]\left(1 + \frac{\Omega}{\Phi} \right) \ln
\varepsilon  \right\} /\ln \varepsilon\, .
\label{4.1.1.29a}
\end{eqnarray}
The right-hand side is non-trivial only when
\begin{eqnarray}
\left|q\left(1 + \frac{\Omega}{\Phi} \right)\right| \ < \ \frac{1}{\sqrt{- \ln
\varepsilon}}\, .
\label{4.1.1.30a}
\end{eqnarray}
[note that $|a_i - \langle a \rangle_q| \sim 1/\sqrt{- \ln \varepsilon}$, see
Appendix ~C]. In such a case  Eq.~(\ref{4.1.1.29a}) can be recast in the
form
\begin{eqnarray}
\tau(q) + a_i (1-q) \ \sim \  q\left(1 + \frac{\Omega}{\Phi} \right) \left[a_i -
\langle a
\rangle_q(\varepsilon) \right]\, ,
\end{eqnarray}
implying that $\Omega/|\Phi| = (2q-1)/q$. With the help of (\ref{4.1.1.30a})
this means that $q \in [1-1/\sqrt{- \ln \varepsilon}, 1 + 1/\sqrt{- \ln
\varepsilon}]$. Bearing this in mind we cab write the single-{\em cell}
probability $p_i \sim \varepsilon^{a_i}$ as
\begin{eqnarray}
p_i \ \sim \   \left[1 + (1-q) (a_i -
\langle a
\rangle_q)\ln
\varepsilon  \right]^{1/(1-q)}\, .
\end{eqnarray}

In multifractals it is more customary to consider the total
probability of a phenomenon with a scaling exponent $a_i$, i.e., $P_i(a) \sim
\varepsilon^{-f(a_i) + a_i}$. To this end we can first utilize a simple
quadratic expansion
\begin{eqnarray}
f(a_i) - f(\langle a \rangle_q) \ = \ q(a_i - \langle a \rangle_q) \ + \
\frac{1}{2}f''(\langle a \rangle_q)(a_i - \langle a \rangle_q)^2  \ + \ \cdots
\ = \
q(a_i - \langle a \rangle_q) \ + \ \frac{1}{2} \frac{(a_i -
\langle a \rangle_q)^2 }{(\Delta a)^2 \ln \varepsilon}\ + \ \cdots \, .
\label{30ac}
\end{eqnarray}
In the last equality we have employed Eqs.~(\ref{C7a})--(\ref{C7b}). Note
also that the higher-order terms in the expansion (\ref{30ac}) are of the order
$\mathcal{O}((-\ln \varepsilon)^{-3/2})$.
From (\ref{M1}) and (\ref{30ac}) we then get
\begin{eqnarray}
P_i^{(1-q)}(a) \ \propto \ 1 +  \ q \left[a_i - \langle a
\rangle_q \right]\left(1 + \frac{\Omega}{\Phi} \right) \ln
\varepsilon  - (1-q) q \left[a_i - \langle a
\rangle_q\right]\ln\varepsilon -(1-q)\frac{1}{2} \frac{(a_i -
\langle a \rangle_q)^2 }{(\Delta a)^2}\, .
\label{32bbc}
\end{eqnarray}
Since for
values $a_i$ close to $\langle a \rangle_q$  the distribution $P_i$ must acquire
(due to curdling theorem) a non-trivial value
in the limit $\varepsilon \rightarrow 0$, the logarithmic divergences in
(\ref{32bbc}) must
cancel each other, yielding the simple condition  $\Omega \ = \ q|\Phi|$ .
With this we can finally write
\begin{eqnarray}
P_i \ \propto \ \left[ 1 \ - \ (1-q) \frac{(a_i -
\langle a \rangle_q)^2 }{2(\Delta a)^2}\right]^{1/(1-q)}\, .
\end{eqnarray}
This distribution is encountered in a number of multifractal systems. A
paradigmatic example can be found in a statistical description of the
intermittent evolution of fully-developed turbulence.
In such a case $P_i(a)$ describes the distribution of singularity exponents of
the velocity gradient~\cite{ArAr1}. In addition, the parameter $q$ satisfies
the scaling relation
\begin{eqnarray}
1/(1-q) \ = \ 1/a_- \ - \ 1/a_+\,,
\end{eqnarray}
where $a_{\pm}$ are defined by $f(a_{\pm}) = 0$. Such a scaling is a
manifestation of the mixing property. In Ref.~\cite{ArAr1} it was further
shown that the $q$ variance $(\Delta a)^2$ can be related to the
phenomenologically important intermittency exponent $\mu$.

%
%
% the rate of convergence is given by the fact that $\Delta a_i(q,\varepsilon) =  a_i(\varepsilon) - \langle a \rangle_q(\varepsilon)$ has to converge at least as ${1}/{\ln \varepsilon}$.
%
%%
%\begin{equation}
%\label{M30}
%\Omega_{r=q} \ \sim \ \frac{\varepsilon^{(q-1)\langle a \rangle_q}(2q -1)}{q}\, .
%\end{equation}
%%
%In addition, (\ref{M30}) together with (\ref{M29}) imply that $\tau = - \langle
%a \rangle_q + q\langle a \rangle_q $ which is equivalent to $\langle a \rangle_q
%= f(\langle a \rangle_q)$.****

%that the underlying statistics describing the intermittent evolution of fully-developed turbulence
%is the one given by

%%%%%%%%%%%%%%%%%%%%%%%%%%%%%%%%%%%%%%%%%%%%%%%%%%%%%%%%%%%%%%
\subsubsection{``High-temperature" expansion}\label{4.1.2a}
%%%%%%%%%%%%%%%%%%%%%%%%%%%%%%%%%%%%%%%%%%%%%%%%%%%%%%%%%%%%%%

\vspace{2mm}
Let us now make an important remark concerning the asymptotic
behavior of $p_i$ in regard to $\Omega$. If we assume that the
Lagrange multiplier $\Omega \ll 1$ then from (\ref{V3b}) the
following expansion holds
\begin{eqnarray}
&&W\!\left( -\frac{\kappa(q-1)}{\Phi q} \ \exp
\left(\frac{q-1}{q}\right) \ \exp\left(-(q-1)\frac{\Omega}{\Phi}
\Delta_q E_i\right) \right) \ \approx \ W\!\left( x \right) \left[
1 \ - \ (1-q) \Omega^* \Delta_q E_i \right]\, ,
\end{eqnarray}
with
\begin{equation}
\Delta_q  E_i \ = \ E_i - \langle E\rangle_q\, , \;\;\;\;\; \Omega^*
\ = \ - \frac{\Omega}{ \Phi(W(x)+1) }\, ,\;\;\;\;\; x \ = \
-\frac{\kappa(q-1)}{\Phi q} \exp\left(\frac{q-1}{q}\right) \, .
\end{equation}
Hence, if we use the relation (\ref{V2}) we can write
\begin{equation}
p_i  \ = \ \frac{\left[1 \ - \ (1-q) \Omega^* \Delta_q E_i
\right]^{1/(1-q)}}{\sum_k \left[ 1 \ - \ (1-q) \Omega^* \Delta_q
E_k \
 \  \right]^{1/(q-1)}} \ = \ Z^{-1} \left[1 \ - \ (1-q) \Omega^*
\Delta_q E_i \right]^{1/(1-q)} \; , \label{Va2}
\end{equation}
with
\begin{eqnarray}
Z = \sum_k \left[ 1 \ - \ (1-q) \Omega^* \Delta_q E_k  \
\right]^{1/(1-q)} = \left[ \frac{q}{\kappa (q-1)}
W(x)\right]^{1/(q-1)}\, .
\end{eqnarray}
The distribution (\ref{Va2}) agrees with the so called 3rd version
of thermostatics introduced by Tsallis {\em et al.}~\cite{Ts2}. It
might by also formally identified with the maximizer for
R\'{e}nyi's entropy~\cite{Ba1}. Clearly, $\Omega^*$ is not a
Lagrange multiplier, but $\Omega^*$ passes to $\beta$  at
$q\rightarrow1$ (in fact, $\Phi \rightarrow -1$, $\Omega
\rightarrow \beta $ and $W(x) \rightarrow 0$ at $q\rightarrow1$).
Note also that when $\Omega =0$ (i.e., no energy constraint) then
$p_i =1/n$ which reconfirms the fact that ${\mathcal{D}}_q$
attains its largest value for the uniform distribution.

%%%%%%%%%%%%%%%%%%%%%%%%%%%%%%%%%%%%%%%%%%%%%%%%%%%%%%%%%%%%%%
\subsubsection{``Low-temperature" expansion}\label{4.1.3a}
%%%%%%%%%%%%%%%%%%%%%%%%%%%%%%%%%%%%%%%%%%%%%%%%%%%%%%%%%%%%%%

\begin{figure}[t]
\begin{center}
\includegraphics[width=8cm]{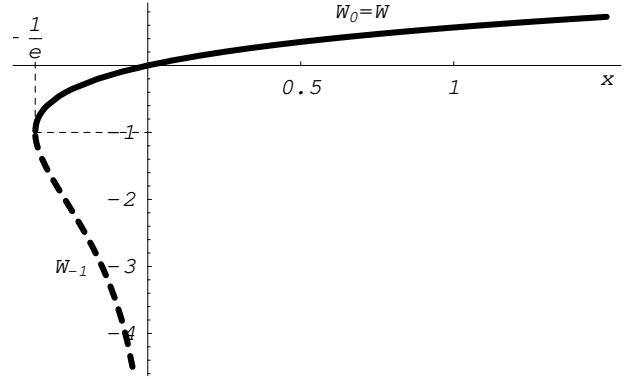}
\caption{Two real branches of the Lambert--W
function. Solid line: $W_{0}(x) \equiv W(x)$ defined for $-1/e
\leq x < +\infty$ (so far considered). Dashed line: $W_{-1}(x)$
defined for $-1/e \leq x < 0$. The two branches meet at point
$(-1/e, -1)$. }
\label{fig4}
\end{center}
\end{figure}

\vspace{2mm}
From the physical standpoint it is the asymptotic behavior at
$\Omega \gg 1$ (or more precisely at $\Omega|(q-1)/\Phi| \gg 1$),
i.e., ``low-temperature" expansion, that is most intriguing. This
is because the branching properties of the Lambert--W function at
negative argument values make the structure of ${\mathcal{P}}$
rather non-trivial. We thus split our task into four distinct
cases:
\begin{eqnarray*}
a_1)\;\;(q-1) > 0 \;\;\;\; &\mbox{and}& \;\;\;\; \Delta_q E < 0\, , \\
a_2)\;\;(q-1) > 0 \;\;\;\; &\mbox{and}& \;\;\;\; \Delta_q E \geq 0\, , \\
b_1)\;\;(q-1) < 0 \;\;\;\; &\mbox{and}& \;\;\;\; \Delta_q E < 0\, , \\
b_2)\;\;(q-1) < 0 \;\;\;\; &\mbox{and}& \;\;\;\; \Delta_q E \geq
0\, .
\end{eqnarray*}
Cases $a_1)$ and $a_2)$ are much simpler to start with as the
argument of $W$ is positive. $W$ is then a real and single valued
function which belongs to the principal branch of $W_0$, see
Fig.\ref{fig4}.
%
%\begin{figure}
%\vspace{4mm} \epsfxsize=8cm \centerline{\epsffile{f1.eps}}
%\vspace{4mm} \caption{Two real branches of the Lambert $W$
%function. Solid line: $W_{0}(x) \equiv W(x)$ defined for $-1/e
%\leq x < +\infty$ (so far considered). Dashed line: $W_{-1}(x)$
%defined for $-1/e \leq x < 0$. The two branches meet at point
%$(-1/e, -1)$. } \label{fig4}
%%\begin{picture}(20,7)
%%\put(200,80){ $q$ } \put(80,55){ $p$ } \put(230,155){$\varrho$}
%%\end{picture}
%\end{figure}
%
When $\Delta_q E < 0$ then $a_1)$ implies the asymptotic
expansion
\begin{eqnarray}
W\left( z \right) \ \approx \ z \;\;\;\; \Rightarrow \;\;\;\; p_i
\ = \ \left(\frac{1}{|\Phi |}\right)^{1/(1-q)}\!e^{-1/q}
\exp\left(-\frac{\Omega}{|\Phi|}\Delta_q E_i\right) \ \equiv \
Z_1^{-1} \exp\left(-\frac{\Omega}{|\Phi|}\Delta_q E_i\right)\, ,
\label{disa}
\end{eqnarray}
with
\begin{eqnarray}
Z_1\ = \
%\ = \ \sum_{k; \Delta_q E_k < 0}
%\exp\left(-\frac{\Omega}{|\Phi|}\Delta_q E_i\right) \ = \
\left(\frac{1}{|\Phi |}\right)^{1/(q-1)}\!e^{1/q}\, .
\end{eqnarray}
Note that in this case $p_i$ is of a Boltzmann type ($\langle E
\rangle_q$ can be canceled against the same term in $Z_1$).

\begin{figure}[t]
\begin{center}
\includegraphics[width=8cm]{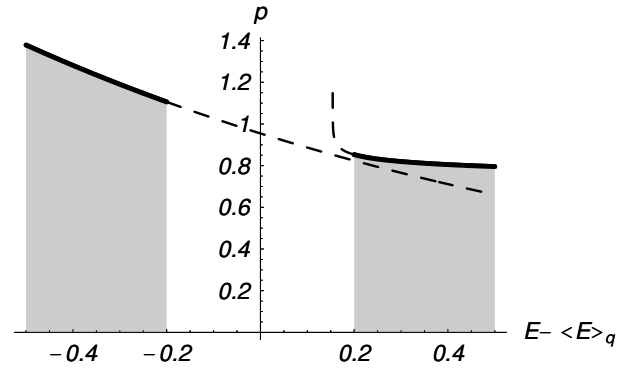}
\caption{A plot of the ``low--temperature" MaxEnt
distribution (\ref{disa})--(\ref{disb}).  The parameters of the
plot are chosen in the following way: $\kappa = 0.01$, $\Phi = -
0.68$, $q = 30$ and $\Omega = 0.5$. The distribution is normalized
to $1$ on the interval ${\Delta}_q E \in [-0.5,0.5]$.}
\label{fig3}
\end{center}
\end{figure}

On the other hand, $a_2)$ situation implies the asymptotic
expansion~\cite{CGHJK}
\begin{eqnarray}
W(z)\ \approx \ \ln(z) - \ln(\ln(z)) \;\;\; \Rightarrow \;\;\;
p_i \ = \ Z_2^{-1} \left[1 - (1-q)\Omega^* \Delta_q E_i
\right]^{1/(1-q)}\, , \label{disb}
\end{eqnarray}
with
\begin{eqnarray}
&&Z_2
%\ = \ \sum_{k; \Delta_q E_k > 0}\left[1 - (1-q)\Omega^*
%\Delta_q E_k \right]^{1/(1-q)}
\ = \ \left[\frac{q}{\kappa (q-1)} \ln\!\left( \frac{\kappa
(q-1)}{|\Phi| q} \exp\left( \frac{q-1}{q} \right)
\right)\right]^{1/(q-1)}\; \; ; \; \; \Omega^* \ = \
\frac{\Omega}{|\Phi|} \left[\ln\!\left( \frac{\kappa (q-1)}{|\Phi| q}
\exp\left( \frac{q-1}{q} \right) \right)\right]^{-1}\, .
\end{eqnarray}
Although the distribution (\ref{disb}) formally agrees with
Tsallis {\em et al.} distribution it cannot be identified with it
as $\Omega^*$ does not tend to $\beta$ in $q\rightarrow1$ limit.
In fact, the limit $q\rightarrow1$ is prohibited in this case as
it violates the ``low-temperature" condition $\Omega|(q-1)/\Phi|
\gg 1$. Note particularly that our MaxEnt distribution represents
in the ``low-temperature" regime a heavy tailed distribution with
Boltzmannian outset. When $\Omega$ and $q
> 1$ are fixed one may find $\kappa$ and $\Phi$ from the
normalization condition
\begin{eqnarray}
Z_1^{-1} \!\!\sum_{k; \Delta_q E_k <
0}\exp\left(-\frac{\Omega}{|\Phi|}\Delta_q E_k\right) \ + \
Z_2^{-1} \!\!\sum_{k; \Delta_q E_k \geq 0}\left[1 - (1-q)\Omega^*
\Delta_q E_k \right]^{1/(1-q)} \ = \ 1\, ,
\end{eqnarray}
and sewing condition at $\Delta_q E = 0$. However, because the
``low-temperature" approximation does not allow to probe regions
with small $\Delta_q E$ one must numerically optimize the sewing
by interpolating the forbidden parts of $\Delta_q E$ axis. Example
of such a numerical optimization is presented in Fig.~\ref{fig3}

%\begin{figure}
%\vspace{4mm} \epsfxsize=8cm \centerline{\epsffile{gy4.eps}}
%\vspace{4mm} \caption{A plot of the ``low--temperature" MaxEnt
%distribution (\ref{disa})--(\ref{disb}).  The parameters of the
%plot are chosen in the following way: $\kappa = 0.01$, $\Phi = -
%0.68$, $q = 30$ and $\Omega = 0.5$. The distribution is normalized
%to $1$ on the interval ${\Delta}_q E \in [-0.5,0.5]$.}
%\label{fig3}
%%\begin{picture}(20,7)
%%\put(200,80){ $q$ } \put(80,55){ $p$ } \put(230,155){$\varrho$}
%%\end{picture}
%\end{figure}

%To eliminate $\kappa$ from (\ref{V2}) we use the following trick. We
%use the definition $\kappa = \sum_k (p_k)^q$ together with the
%relation (\ref{VIIIc}), then
%
%\begin{eqnarray}
%&&\frac{d\kappa(q)}{dq} \ = \ \sum_k (p_k)^q \ln p_k \ = \ \kappa(q)
%\
%\frac{\ln(-\Phi)}{q-1}\nonumber \\
%&&\frac{d\Phi(q)}{dq} \ = \
%\end{eqnarray}
%

Cases $b_1)$ and $b_2)$ are technically more involved, because
$q < 1$ causes that the argument of $W(\cdots)$ is negative. In case
$b_1)$ we obtain for low temperatures that
\begin{eqnarray}
\frac{\kappa (q-1)}{q} \exp\left(-(q-1)
\frac{\Omega}{\Phi} \Delta_q E_i\right) \  \rightarrow \  -\infty\, .
\end{eqnarray}
Nevertheless, the
complex Lambert--W function has a branch cut in the interval
$\left[-\infty,-{1}/{e}\right]$, so the real-valued Lambert--W function is
defined only for $x > -{1}/{e}$ and Eq.~\eqref{V3b} has no real solution. This
situation corresponds previous discussions about existence of solution of
Eq.~\eqref{V4}.

In case $b_2)$ there exist two solutions of Eq.~\eqref{V3b}, i.e. $W_0(z)$ and
$W_{-1}(z)$ (see Fig.~\ref{fig4}). In case of the principal branch $W_0(z)$, the
Lambert W--function can be approximated as $W(z) \approx z$, and the solution
corresponds to the case $a_1)$. In case of the principal branch $W_{-1}(z)$, the
asymptotic expansion for $z \rightarrow 0^{-}$ is
\begin{equation}
W_{-1}(z) \ \! \approx  \ \! \ln (-z) - \ln (- \ln (-z))
\end{equation}
so the resulting probability is similar to the case $a_2)$, only with
\begin{eqnarray}
\Omega^* \ = \ \frac{\Omega}{|\Phi| } \left[\ln \left( \frac{\kappa (1-q)}{|\Phi| q}
\exp\left( \frac{q-1}{q} \right) \right)  \right]^{-1}\, .
\end{eqnarray}

We should stress that for all cases it is necessary to check the validity of the
asymptotic
expansion and its applicability to the MaxEnt distribution. In some cases can
the expansion violate the condition $p_i \leq 1$ and then it is not possible to
use such approximations.

%%%%%%%%%%%%%%%%%%%%%%%%%%%%%%%%%%%%%%%%%%%%%%%%%%%%%%%%%%%%%%
\subsection{ The $r=1$ case}
%%%%%%%%%%%%%%%%%%%%%%%%%%%%%%%%%%%%%%%%%%%%%%%%%%%%%%%%%%%%%%

\vspace{2mm}
When $r = 1$ is chosen (i.e., when the constraints are implemented
via the usual linear averaging) then Eq.(\ref{VIIIa}) implies
\begin{eqnarray}
\Phi  \ = \ e^{(q-1){\langle \ln{\mathcal{P}} \rangle_q}}
\left[q\left({\langle \ln{\mathcal{P}} \rangle_q} - \ln p_i\right)
- 1\right]\, \frac{(p_i)^{q-1}}{\sum_k (p_k)^q} \ - \ \Omega (E_i
- \langle E\rangle ) \, . \label{V5}
\end{eqnarray}
Multiplying by $p_i$ and summing over $i$ we obtain the constraint
\begin{eqnarray}
\Phi \ = \ -  e^{(q-1)\langle \ln {\mathcal{P}} \rangle_q } \, \,
\Rightarrow \, \, \frac{\ln(-\Phi)}{q-1} \ = \ \langle \ln
{\mathcal{P}} \rangle_q \, . \label{V6}
\end{eqnarray}

%\vspace{0mm}

%\noi Resolving (\ref{V6}) for  $\langle \ln{\mathcal{P}}
%\rangle_q$ we find
%
%\begin{eqnarray}
%&&\langle \ln{\mathcal{P}} \rangle_q = -{\mathcal{H}} + 1/q +
%\frac{W\left( x\right)}{(q-1)}\, , \nonumber \\
%&& x = e^{(q-1)({\mathcal{H}} - 1/q)} \ \frac{(q-1)}{q} \Phi\, .
% \label{V6b}
%\end{eqnarray}
\noi Upon insertion of (\ref{V6}) into (\ref{V5}) we get a
transcendental equation for $p_i$, which reads
\begin{equation}\label{eq: r1}
\kappa p_i^{1-q} \ = \ \frac{\Phi}{\Phi+\Omega (E_i - \langle E \rangle)}\left[q \ln p_i - \frac{q \ln(-\Phi)}{q-1} + 1\right]\, .
\end{equation}
The solution can be again written in terms of the Lambert W--function, namely
\begin{eqnarray}
 p_i \ &=& \
     \left[\frac{q\Phi}{(q-1) \kappa (\Phi + \Omega \Delta E_i)}
     \;\;
             W\!\left( -\frac{ \kappa(q-1)}{{\Phi }q} \
\exp\left(\frac{q-1}{q}\right)\ \left(1 +  \frac{\Omega}{\Phi}
\Delta E_i \right) \right)
            \right]^{1/(1 - q)}\nonumber \\[3mm]
            &=& \ \frac{1}{(- \Phi)^{1/(1-q)} e^{1/q}}\
             \exp\left\{ \frac{1}{(q-1)}\ W\!\left( -\frac{ \kappa(q-1)}{{\Phi }q} \
\exp\left(\frac{q-1}{q}\right)\ \left(1 +  \frac{\Omega}{\Phi}
\Delta E_i \right) \right)
                    \right\}\, .
                    \label{V7}\\[1mm]\nonumber
\end{eqnarray}
Relations (\ref{V3a}) nad (\ref{V3b}) again ensure that all
$p_i$'s are positive. In addition, it is easy to check that in the
limit case $q\rightarrow1$, the formula (\ref{V7}) approaches the
classical Gibbsian maximizer. Indeed, if we utilize the
identities:
\begin{eqnarray}
\kappa|_{q=1} \ = \ 1, \; \Phi|_{q=1} \ = \ -1, \;
\left[(-\Phi)^{1/(1-q)}\right]|_{q=1} \ = \ e^{-{\mathcal{H}}}, \;\;
\mbox{and} \;\; \Omega|_{q=1} \ = \ \beta\, , \label{IVB}
\end{eqnarray}
then
\begin{eqnarray}
p_i|_{q=1} \ = \ e^{-{\mathcal{H}} \ + \ \beta (\langle E \rangle -
E_i)} \ = \ e^{\beta F - \beta E_i} \ = \ e^{-\beta E_i}/Z\, .
\end{eqnarray}
Similarly as Eq.(\ref{IVA44}) also the relation (\ref{IVB})
represents an important consistency check of our procedure.

%To eliminate
%$\kappa$ from (\ref{V7}) we again use the fact that

%%%%%%%%%%%%%%%%%%%%%%%%%%%%%%%%%%%%%%%%%%%%%%%%%%%%%%%%%%%%%%
\subsubsection{Multifractal case}
%%%%%%%%%%%%%%%%%%%%%%%%%%%%%%%%%%%%%%%%%%%%%%%%%%%%%%%%%%%%%%

\vspace{2mm}
By following the same strategy as in the case $r=q$, we plug the multifractal scaling relations for $p_i \sim \varepsilon^{a_i}$ to Eq.~\eqref{V6}
and use the fact that
the role of $E_i$ is taken over by $-a_i\ln \varepsilon$. After a short calculation we arrive at
\begin{eqnarray}
&&\varepsilon^{\tau(q) + a_i(1-q)} \left[ 1 - \frac{\Omega}{\Phi}\left(a_i -
\langle a \rangle_1\right) \ln \varepsilon \right] \ \sim \  1 \ + \ q\left(a_i
- \langle a \rangle_q \right) \ln \varepsilon\, ,
% &&\varepsilon^{\tau} - \varepsilon^{a_i (q-1)} \ \sim \ \left[\Delta a_i(q,\varepsilon) q \varepsilon^{a_i (q-1)} \ + \ \frac{\Omega}{\Phi} \varepsilon^\tau \Delta a_i(1,\varepsilon)\right] \ln \varepsilon\, .
\end{eqnarray}
which in the small-$\varepsilon$ limit yields
\begin{eqnarray}
\tau(q) + a_i(1-q) \ + \ \ln\left[1 - \frac{\Omega}{\Phi}\left(a_i -
\langle a \rangle_1\right) \ln \varepsilon   \right]/\ln \varepsilon \ \sim \
\ln\left[1 + q\left(a_i -
\langle a \rangle_q\right) \ln \varepsilon   \right]/\ln \varepsilon \ \sim \
0\, .
\label{4.2.1.56a}
\end{eqnarray}
The last relation follows from the fact that for $q\geq 1/2$ (cf. \ref{AppA})
the expression goes to zero in the small $\varepsilon$ limit.
Note that Eq.~(\ref{4.2.1.56a}) implies a nontrivial behavior only for
\begin{eqnarray}
{\Omega}/{|\Phi|} \ < \ {1}/{\sqrt{-\ln \varepsilon}}\, .
\end{eqnarray}
This then gives that
\begin{eqnarray}
\tau(q) + a_i(1-q) \ \sim \ \frac{\Omega}{\Phi} (a_i -
\langle a \rangle_1)\, ,
\label{4.2.1.58a}
\end{eqnarray}
and hence $\Omega/|\Phi| = q-1$. Latter shows in particular that $q \in
[1, 1+ 1/\sqrt{-\ln \varepsilon}]$. Rather than dealing with the single-{\em
cell} probability $p_i$ we can again address the (more relevant)
total probability $P_i(a)\sim \varepsilon^{a_i - f(a_i)}$. By using the fact
that (cf. Eq.~(\ref{4.2.1.58a}))
\begin{eqnarray}
\tau(q) + [a_i-f(a_i)](1-q) + \frac{\Omega}{|\Phi|}(a_i -
\langle a \rangle_1) + f(a_i)(1-q) \ \sim \ 0,
\end{eqnarray}
and
% %
% \begin{eqnarray}
% P_i^{(1-q)}(a) \ \sim \ \left[ 1 \ + \ q\left(a_i - \langle a \rangle_q \right) \ln \varepsilon \ - \ (1-q) (a_i - \langle a\rangle_1)\ln \varepsilon   \right]/\left[ 1 - \frac{\Omega}{\Phi}\left(a_i - \langle a \rangle_1\right) \ln \varepsilon \right].
% \end{eqnarray}
% %
% Here, we have used
the expansion
\begin{eqnarray}
f(a_i) - f(\langle a \rangle_1) \ = \ (a_i - \langle a \rangle_1) \ + \
\frac{1}{2}f''(\langle a \rangle_1)\left(a_i - \langle a \rangle_1\right)^2 \ + \ \cdots  \ = \
\left(a_i - \langle a \rangle_1\right) \ + \ \frac{1}{2} \frac{(a_i -
\langle a \rangle_1)^2 }{(\Delta a)^2 \ln \varepsilon} \ + \ \cdots  \, .
\label{4.2.1.54.aa}
\end{eqnarray}
[in the second equality  we have used  again the curdling
theorem (see \ref{AppC})], we obtain
% The right-hand side of the relation gives again the convergence to the prevailing exponent. We get two interesting results. Firstly $\Delta a_i(q,\varepsilon)$ and $\Delta a_i(1,\varepsilon)$ have to tend to zero at least as ${\varepsilon^{-\tau(q)}}/{\ln \varepsilon}$, and
%
\begin{eqnarray}
P_i(a) \ \sim \ \left[ 1 \ - \ \frac{(1-q)}{2} \frac{\left(a_i - \langle
a \rangle_1\right)^2}{(\Delta a)^2} \right]^{1/(1-q)}.
\end{eqnarray}
%
% \begin{equation}
% \Omega_{r=1} \ \sim \ q \Phi \frac{\Delta a_i(q,\varepsilon)}{\Delta a_i(1,\varepsilon)} \ \approx \ q \Phi \ = \ q \Omega_{r=q}\, .
% \end{equation}
This prescription naturally appears in the context of multiplicative cascades
with the coarse-grained scaling $\varepsilon = 2^{-k}$ ($k\geqq 1$).  Again, the
natural application would be in a fully-developed turbulence. The proximity of
$q$ to one makes the previous distribution suitable for discussions concerning
the dynamics on the {\em measure theoretic support}, i.e., a set whose
Hausdorff--Besicovich dimension is $a(1) = f(a(1))$. In particular, it can be
shown~\cite{Fed1} that the measure theoretic support describe the set on which
the probability is concentrated.

%%%%%%%%%%%%%%%%%%%%%%%%%%%%%%%%%%%%%%%%%%%%%%%%%%%%%%%%%%%%%%
\subsubsection{``High-temperature" expansion}
%%%%%%%%%%%%%%%%%%%%%%%%%%%%%%%%%%%%%%%%%%%%%%%%%%%%%%%%%%%%%%

\vspace{2mm}
Similarly as in the $r=q$ case we can find the ``high-temperature"
expansion by assuming that $\Omega \ll 1$. In such a case we have
\begin{eqnarray}
\frac{\Phi}{(\Phi + \Omega \Delta E_i)} \; W\!\left( -\frac{
\kappa(q-1)}{{\Phi }q} \ \exp\left(\frac{q-1}{q}\right)\ \left(1 +
 \frac{\Omega}{\Phi} \Delta E_i \right) \right) \ \approx \
W(x)\left[ 1  -  (1-q) \Omega^* \Delta E_i \right]\, .
\end{eqnarray}
Here
\begin{eqnarray}
\Omega^* = - \frac{q}{q-1}\frac{ \Omega W(x)}{\Phi (W(x) +1)}\, ,
\;\;\;\;\; x \ = \ - \frac{\kappa (q-1)}{\Phi q}
\exp\left(\frac{q-1}{q} \right)\, .
\end{eqnarray}
Through (\ref{V7}) this implies that
\begin{equation}
p_i  \ = \ Z^{-1} \left[1 \ - \ (1-q) \Omega^* \Delta E_i
\right]^{1/(1-q)} \; , \label{Vb2}
\end{equation}
with
\begin{eqnarray}
Z = \sum_k \left[ 1 \ - \ (1-q) \Omega^* \Delta E_k  \
\right]^{1/(1-q)} = \left[ \frac{q}{\kappa (q-1)}
W(x)\right]^{1/(q-1)}\, .
\end{eqnarray}
Relation (\ref{Vb2})  coincides with the Tsallis-type distribution that is historically known as Bashkirov's 1st version of thermostatistics~\cite{Ba1}.

Note in passing that by using the identity $\lim_{q\rightarrow1} W(x)/(q-1) =1$,
we obtain that the factor $\Omega^*$ approaches
the inverse temperature $\beta$  in the limit $q\rightarrow 1$ as it should.

%%%%%%%%%%%%%%%%%%%%%%%%%%%%%%%%%%%%%%%%%%%%%%%%%%%%%%%%%%%%%%
\subsubsection{``Low-temperature" expansion}
%%%%%%%%%%%%%%%%%%%%%%%%%%%%%%%%%%%%%%%%%%%%%%%%%%%%%%%%%%%%%%

\vspace{2mm}
We now wish to consider the ``low-temperature" expansion --- i.e.,
$\Omega \gg 1$. Similarly as in the $r=q$ case, we divide the situation into four sub-cases:
\begin{eqnarray*}
a_1)\;\;(q-1) > 0 \;\;\;\; &\mbox{and}& \;\;\;\; \Delta E < 0\, , \\
a_2)\;\;(q-1) > 0 \;\;\;\; &\mbox{and}& \;\;\;\; \Delta E \geq 0\, , \\
b_1)\;\;(q-1) < 0 \;\;\;\; &\mbox{and}& \;\;\;\; \Delta E < 0\, , \\
b_2)\;\;(q-1) < 0 \;\;\;\; &\mbox{and}& \;\;\;\; \Delta E \geq
0\, .
\end{eqnarray*}
Unlike case $r=q$, the sub-cases group into two qualitatively distinct classes:
\begin{enumerate}
  \item cases $a_2)$ and $b_1)$ lead to the asymptotic expansion $W(z) \propto \ln (z) - \ln( \ln (z))$, because $- \frac{k(q-1)\Delta E_i}{\Phi^2 q} > 0$.
  \item cases $a_1)$ and $b_2)$ lead to the situation, when the Lambert W--function is not defined, which corresponds to the fact that Eq.~\eqref{eq: r1} has no solution.
\end{enumerate}
So in particular, we see that in cases when our hybrid entropy cannot be consistently used over the whole temperature range. It can be
at best used as an effective entropy in higher-temperature regimes. This might be particularly pertinent in the high-energy particle
phenomenology where the host of phase transitions is happening under conditions that are far from thermal equilibrium (e.g., chiral phase transition in QCD and ensuing quark-gluon plasma formation). In the first case, i.e., when the asymptotic expansion exists, the probability distribution can be written in the form
\begin{equation}
p_i\ = \ \left\{ \frac{ \frac{\kappa (q-1)}{q} \left(1+ \Omega \Delta E_i \right)}{\ln \left[ - \frac{\kappa (q-1)}{q \phi}\exp\left(\frac{q-1}{q}\right)\right] + \ln\left[1 + \Omega \Delta E_i \right]}\right\}^{{1}/(q-1)}\, .
\end{equation}
Contrary to $r=q$, the resulting distribution has functionally different form from both the Boltzmann distribution and Tsallis distribution, even in the generalized form, i.e. with the self-referential temperature. For large temperatures, the second term in the denominator is negligible and the distribution becomes similar to power-like behavior. We shall again note that it is necessary to check consistency of asymptotic expansions.

%%%%%%%%%%%%%%%%%%%%%%%%%%%%%%%%%%%%%%%%%%%%%%%%%%%%%%%%%%%%%%%%%%%%%%%%%%%%%%%%%%%%%%
\section{Concavity and Schur-concavity of ${\mathcal{D}}_q$ } \label{sec: concavity}
%%%%%%%%%%%%%%%%%%%%%%%%%%%%%%%%%%%%%%%%%%%%%%%%%%%%%%%%%%%%%%%%%%%%%%%%%%%%%%%%%%%%%
%

\vspace{2mm}

In this section we discuss the concavity properties of ${\mathcal{D}}_q$. When referring to concavity issue of entropies, one has to distinguish
between two types of concavity. In thermodynamics, the important issue is to show whether or not  the thermodynamical entropy is a concave function of extensive variables. This means to show that ${\mathcal{D}}_q|_{\mbox{\footnotesize{max}}}$ is a concave function under the constraints as in the case of Gibbsian MaxEnt.
%In other words we wish to show that
%
%\begin{eqnarray}
%{\mathcal{D}}_q(\lambda \langle E^{(1)} \rangle_r +
%(1-\lambda)\langle E^{(2)} \rangle_r
%)|_{\mbox{\footnotesize{max}}} \ \geq \ \lambda
%{\mathcal{D}}_q(\langle E^{(1)}
%\rangle_r)|_{\mbox{\footnotesize{max}}} +
%(1-\lambda){\mathcal{D}}_q(\langle E^{(2)}
%\rangle_r)|_{\mbox{\footnotesize{max}}}\, .
%\end{eqnarray}
%
%for any $\lambda \in [0,1]$ and $r=1$ or $r=q$.
Note that in contrast to the information-theoretic entropy
${\mathcal{D}}_q$, ${\mathcal{D}}_q|_{\mbox{\footnotesize{max}}}$
is the system entropy, i.e., it depends on the actual system state
variables.

In the information theory, the significance of concavity lies in the fact that it automatically ensures the validity of the maximality axiom. In case of ${\mathcal{D}}_q$, it suffices to explore the concavity issue only for $e^{- \langle \ln \mathcal{P} \rangle_q}$ because $\ln_{\{q\}}$ is concave and non-decreasing function for all $q>0$,
\begin{equation}
\frac{\partial^2}{\partial p_i^2} \left(e^{- \langle \ln \mathcal{P} \rangle_q} \right) \ = \ e^{- \langle \ln \mathcal{P} \rangle_q} \left[\left(\frac{\partial \langle \ln \mathcal{P} \rangle_q}{\partial p_i}\right)^2 - \frac{\partial^2 \langle \ln \mathcal{P} \rangle_q}{\partial p_i^2} \right].
\end{equation}
It can be shown that the bracket is always negative for $q \in [\frac{1}{2},1]$. Contrary, for $q>1$ we have that $\frac{\partial^2 \langle \ln \mathcal{P} \rangle_q}{\partial p_i^2}\left|_{p_i \rightarrow 0}\right. = - \infty$, while the first term remains bounded and therefore the function cannot be concave for all $p_i$'s.

However, concavity is only a sufficient condition that ensures the maximality axiom. As known, e.g., from the case of R\'{e}nyi entropy, there are examples of non-concave entropies which still have well defined global maximum at $\mathcal{P} = \{1/n, \ldots, 1/n\}$. In fact, there exist weaker concepts
that ensure validity of the maximality axiom. Among these the most prominently is the notion of Schur-concavity~\cite{Schur}. The overview of applications of Schur-concavity can be found in Refs.~\cite{SC,RV}. This concept is based on the idea of majorization. We say that a probability distribution $\mathcal{P} = \{p_1,\dots,p_n\}$ is majorized by distribution $\mathcal{Q} = \{q_1,\dots,q_n\}$ if for ordered probability vectors $p_{(1)} \geq p_{(2)} \geq \dots$, resp. $q_{(1)} \geq q_{(2)} \geq \dots$ hold $\sum_{k=1}^j p_{(k)} \leq \sum_{k=1}^j q_{(k)}$, where $j=1,\dots,{n-1}$ (for $j=n$ is the inequality fulfilled automatically from normalization). We denote $\mathcal{P} \prec \mathcal{Q}$. We say that the function $F$ is Schur-concave if for $\mathcal{P} \prec  \mathcal{Q}$ is $F(\mathcal{P}) \geq F(\mathcal{Q})$.
The Schur-concavity automatically preserves the maximality axiom, because the uniform distribution is majorized by every other distribution. Shi et al. have shown~\cite{Shi} that special subclass of functions called Gini means (defined e.g. in Ref.~\cite{Gini}) that can be expressed in the form
\begin{equation}
G(q;x,y) = \exp\left(\frac{x^q \ln x + y^q \ln y}{x^q + y^q}\right)\, ,
\end{equation}
is for $(x,y) \in \mathds{R}^2_+$ Schur-convex function of $(x,y)$ when $2q \geq 1$ (this is a consequence of \cite[Theorem 1]{Shi} for $r=s$). It is then easy to shown that $\mathcal{D}_q$ is Schur-concave function for $q \geq \frac{1}{2}$. As a consequence, for $q \geq \frac{1}{2}$, $\mathcal{D}_q$ fulfills the maximality axiom. Moreover, Ref.~\cite{Shi} discussed the case $q \in (0,\frac{1}{2}$ and concluded that one cannot say anything about Schur-convexity or Schur-concavity of ${\mathcal{D}}_q$ on this interval. For illustration, in Fig.~\ref{Fif.3bb} we compare three types of entropies, i.e., $\mathcal{D}_q$, $\mathcal{S}_q$ and $\mathcal{I}_q$ for various values of $q$ on distribution ${\mathcal{P}} = \{p,1-p\}$, and we observe that $\mathcal{D}_{0.4}$ is neither Schur-convex nor Schur-concave, which is caused by the fact that maximum is not in $\mathcal{P} = \left\{1/n,\dots,1/n\right\}$.

%%%%%%%%%%%%%%%%%%%%%%%%%%%%%%%%%%%%%%%%%%%%
\section{Conclusions and outlooks}\label{IX}
%%%%%%%%%%%%%%%%%%%%%%%%%%%%%%%%%%%%%%%%%%%%

We have presented a plausible generalization of the information
entropy concept. Our approach is based on an axiomatic merger of two
currently widely used information measures: R\'{e}nyi's and
Tsallis--Havrda--Charv\'{a}t's. Such a merger is natural from the
mathematical point of view as both above measures have an axiomatic
underpinning with a very similar axiomatics. From the physics
viewpoint the above merger is interesting because it combines two
entropies with analogous MaxEnt distributions but with very
different scope of applicability in physics.
% On mathematical side the
% entropy ${\mathcal{D}}_q$ is closely related with Czis\'{a}r's measure
% of directed divergence~\cite{} ****

We have shown that the maximizers for ${\mathcal{D}}_q$ subject to
constant averaged energy are represented in terms of the Lambert
W--function. The Lambert W--function is a special function that
appears in numerous exactly solvable statistical
systems. Tonks gas~\cite{Caillal}, Richards growth model and Lotka--Volterra models~\cite{pep}
may serve as examples. The Lambert W--function was
recently also used in quantum statistics~\cite{Valluri} and statistics of weak long-range repulsive potentials~\cite{Caillal}. This usage nicely bolsters our suggestion that a typical playground for ${\mathcal{D}}_q$ could be in
statistical systems with both self-similarity and non-locality. In addition, as a byproduct,
we have obtained during our analysis  some new mathematical properties of the Lambert
W--function.

Due to complicated analytical structure of the MaxEnt distribution
we have resorted in our discussion to the
``low" and ``high temperature" asymptotic regimes. We have shown
that under certain parameter conditions these have the heavy
tailed behavior that is identical with Tsallisian maximizers. The
fact that this is true only asymptotically might be at first sight
a bit surprising, as there exists perception that both THC and
R\'{e}nyi's entropies have the same maximizer and hence the merger
entropy should again posses the same MaxEnt distribution. This
anticipation is clearly erroneous. Indeed, both R\'{e}nyi entropy
and THC maximizers have the same {\em functional form} but their
respective``temperature" parameters
$\beta^*$ are entirely different functions of $q$, and in the case
of THC entropy $\beta^*$ is even self-referential (i.e., it
depends on the distribution itself)~\cite{Ba1}.

%Further investigation should address the ***

%By collecting these results ***

In summary, we have shown that there exists a well defined sense in which one can combine
R\'{e}nyi and THC entropic paradigms. We have found the associated one-parametric class of
entropy measures, namely~\eqref{IIIh} and the ensuing MaxEnt
distributions~\eqref{V2}.
It can be rightly objected that apart from the axiomatic side more is needed to consider
${\mathcal{D}}_q$ as a legitimate object of statistical physics.
In this connection one should, however, stress that the presented entropy
has a number of desirable attributes; like THC entropy it is a one-parametric class of
entropies satisfying the non-extensive $q$-additivity, it goes over into ${\mathcal{H}}({\mathcal{P}})$ in the $q\rightarrow 1$ limit,
it complies with thermodynamic stability, continuity, symmetry,
expansivity, decisivity, Schur concavity, etc. On that basis it appears that both
${\mathcal{D}}_q$ and THC entropies have an equal right to serve as
a generalization of statistical thermodynamics.

%  It remains to be seen whether any {\em real} physical
% system can be found to confirm this type of behavior. As for the
% entropy ${\mathcal{D}}_{q}$ itself, it is proper to emphasize that
% apart from presented axiomatic (or constructive) justification,
% the operational characterization of ${\mathcal{D}}_{q}$, similarly
% as ${\mathcal{S}}_q$, would be needed. Work along those lines is
% presently in progress.

%%%%%%%%%%%%%%%%%%%%%%%%%%%%%%%%%%%%%%%%%%%%
\section{Acknowledgements}\label{XI}
%%%%%%%%%%%%%%%%%%%%%%%%%%%%%%%%%%%%%%%%%%%%

We acknowledge very helpful discussions with P.T.~Landsberg, P.~Haramo\"{e}s and
H.~Lavi\v{c}ka which have helped us to understand better the ideas discussed in this paper.
The work was supported by the Grant Agency of the CTU in Prague, grant No. SGS13/217/OHK4/3T/14 and the GA\v{C}R, grant No. GA14-07983S.

\appendix

%%%%%%%%%%%%%%%%%%%%%%%%%%%%%%%%%%%%%%%%%%%%%%%%%%%%%%%%%%%%%%%%%%%%%%
\section{Derivation of ${\mathcal{D}}_q$ from J-A axioms}\label{AppA}
%%%%%%%%%%%%%%%%%%%%%%%%%%%%%%%%%%%%%%%%%%%%%%%%%%%%%%%%%%%%%%%%%%%%%%

In this appendix we show the basic steps in the derivation of functional form
of hybrid entropy ${\mathcal{D}}_q$.

Let us first denote ${\mathcal{D}}(1/n,1/n, \ldots, 1/n) =
{\mathcal{L}}(n)$. Axioms $2$ and $5$ then imply that
${\mathcal{L}}(n) = {\mathcal{D}}(1/n,\ldots, 1/n,0) \ \leq \
{\mathcal{D}}(1/{n+1}, \ldots, 1/{n+1}) = {\mathcal{L}}(n+1)$.
Consequently ${\mathcal{L}}$ is a non-decreasing function of $n$.
To determine the explicit form of ${\mathcal{L}}(n)$ we will
assume that $A^{(1)}, \ldots, A^{(m)}$ are independent experiments
each with $r$ equally probable outcomes, so
\begin{eqnarray}
{\mathcal{D}}(A^{(k)}) = {\mathcal{D}}(1/r, \ldots, 1/r) =
{\mathcal{L}}(r)\, , \;\;\;\; (1 \leq k \leq m)\, .
\end{eqnarray}
Repeated application of axiom $3$ then leads to
\begin{eqnarray}
{\mathcal{D}}(A^{(1)}\cup A^{(2)} \cup \ldots \cup A^{(m)})=
{\mathcal{L}}(r^m) &=& \sum_{k=1}^m {}^mC_k \ (1- q)^{k-1}
{\mathcal{D}}^k(A^{(k)})\nonumber \\
&=& \frac{1}{(1- q)} \left[ \left( 1 + (1 - q){\mathcal{L}}(r)
\right)^m - 1\right]\, , \label{VI1}
\end{eqnarray}
\noindent where $^mC_k$ is the binomial coefficient. By assuming that (\ref{VI1}) can be extended from
$m\in {\mathbb{N}}_+$ to ${\mathbb{R}}_+$ we can take partial
derivative of both sides of (\ref{VI1}) with respect to $m$ and by
setting $m=1$ we obtain the differential equation
\begin{eqnarray}
\frac{(1-q) \ d {\mathcal{L}}}{(1 + (1-q) \
{\mathcal{L}})\left[\ln\left(1 + (1-q)\ {\mathcal{L}}
\right)\right] } \ = \ \frac{dr}{r \ln r}\, . \label{VI2}
\end{eqnarray}
The general solution of (\ref{VI2}) has the form
\begin{eqnarray}
{\mathcal{L}} (r) \ \equiv \ {\mathcal{L}}_q (r) \ = \
\frac{1}{1-q}\left( r^{c(q)} -1\right)
% \ =
%\ \ln_{q}r  \ = \ -\ln_{2-q}\left(\frac{1}{r}\right)
\, . \label{III0a}
\end{eqnarray}
The integration constant $c(q)$ will be determined shortly. Right now we just note
that because at $q = 1$ Eq.(\ref{VI1}) boils down to
${\mathcal{L}}(r^m) = m{\mathcal{L}}(r)$ we must have $c(1) = 0$.
In addition, the monotonicity of ${\mathcal{L}}(r)$ ensures that
$c(q)/(1-q) \geq 0$. To proceed further let us consider the
experiment with outcomes $A = (A_1,A_2, \ldots, A_n)$ and the
distribution ${\mathcal{P}} = \{p_1, p_2, \ldots, p_n\}$. Assume
moreover that $p_k \, (1\leq k \leq n)$ are rational numbers,
i.e.,
\begin{eqnarray}
p_k = \frac{g_k}{g}\, , \;\;\;\; \sum_{k=1}^n g_k = g\, , \;\;\;\;
g_k \in \mathbb{Z}^+\, .
\end{eqnarray}
Let have, in addition, an experiment $B = (B_1, B_2,\ldots, B_g)$
with associated distribution ${\mathcal{Q}} = \{q_1, q_2, \ldots,
q_g \}$. We split $(B_1,B_2,\ldots, B_g)$ into $n$ groups
containing $g_1, g_2, \ldots, g_n$ outcomes respectively. Consider
now a particular situation in which whenever event $A_i$ in $A$
happens then in $B$ all $g_k$ events of $k$-th group occur with
the equal probability $1/g_k$ and all the other events in $B$ have
probability zero. Hence
\begin{eqnarray}
{\mathcal{D}}(B|A = A_k) = {\mathcal{D}}(1/g_k, \ldots, 1/g_k) =
{\mathcal{L}}_q( g_k) \, ,
\end{eqnarray}
\noindent and so axiom $3$ implies that
\begin{eqnarray}
{\mathcal{D}}(B|A) = f^{-1}\left( \sum_{k=1}^n \varrho_k(q)
f({\mathcal{L}}_q (g_k)) \right) \, .
\end{eqnarray}
\noindent On the other hand, in the stated system  the entropy
${\mathcal{D}}(A \cup B)$ can be easily evaluated. Realizing that
the joint probability distribution corresponding to $A \cup B$ is
\begin{eqnarray}
{\mathcal{R}} \ = \ \left\{ r_{kl}\ = \ p_k q_{l|k}
 \right\} \ = \  \{ \underbrace{\frac{p_1}{g_1}, \ldots, \frac{p_1}{g_1}, }_{g_1
\times} \underbrace{\frac{p_2}{g_2}, \ldots,\frac{p_2}{g_2},
}_{g_2 \times} \ldots, \underbrace{\frac{p_n}{g_n}, \ldots,
\frac{p_n}{g_n} }_{g_n \times}\} \ = \ \left\{ 1/g, \ldots, 1/g
\right\}\, ,
\end{eqnarray}
\noindent we obtain that ${\mathcal{D}}(A\cup B) =
{\mathcal{L}}_q(g)$. Applying axiom 3 together with (\ref{III0a})
we get
\begin{eqnarray}
&&{\mathcal{D}}(A) \left( 1+ (1-q) f^{-1}\left( \sum_k
\varrho_k(q) f({\mathcal{L}}_q(p_k) [1 + (1-q) {\mathcal{L}}_q(g)]
+ {\mathcal{L}}_q(g) )
 \right)
 \right)\nonumber \\
&&\mbox{\hspace{2cm}} = \ {\mathcal{L}}_q(g) - f^{-1}\left( \sum_k
\varrho_k(q) f({\mathcal{L}}_q(p_k) [1 + (1-q) {\mathcal{L}}_q(g)]
+ {\mathcal{L}}_q(g) )
 \right)\, .
\end{eqnarray}
\noindent Define $f_{(a,y)}(x) = f(-ax + y)\,  \Rightarrow \
f^{-1}(x) - y = -a f^{-1}_{(a,y)}(x)\;$ then
\begin{eqnarray}
&&{\mathcal{D}}(A) = \frac{f^{-1}_{(a, {\mathcal{L}}(g))} \left(
\sum_k \varrho_k(q) f_{(a, {\mathcal{L}}_q(g))}(-{\mathcal{L}}
(p_k)) \right)}{1-(1-q) f^{-1}_{(a, {\mathcal{L}}(g))} \left(
\sum_k \varrho_k(q) f_{(a, {\mathcal{L}}_q(g))}(-{\mathcal{L}}
(p_k)) \right)}, \label{IIIa}
\end{eqnarray}
\noindent
%By axiom 4 $f(x)$ is invertible in $[0,\infty )$ and so
%$f_{(a, {\mathcal{L}}(g))}(x)$ is invertible in $[0,\infty )$.
with $a = [1 + (1-q) {\mathcal{L}}_q(g)]$.

To proceed further, let us formally put $p_k = 1/r$.
Eq.~(\ref{III0a}) then indicates that it is
${\mathcal{L}}_q(1/p_k)$ and not $-{\mathcal{L}}_q(p_k)$ which
represents the elementary information of order $q$ affiliated with
$p_k$ (cf. with Eq.~(\ref{IIC6})). It is thus convenient to
reformulate (\ref{IIIa}) directly in terms of
${\mathcal{L}}_q(1/p_k)$. This can be done via relation
\begin{eqnarray}
{\mathcal{L}}_q(p_k) =
-\frac{{\mathcal{L}}_q(1/p_k)}{1+(1-q){\mathcal{L}}_q(1/p_k)}\, .
\end{eqnarray}
If we now write
\begin{eqnarray}
{\textsl{g}}(x) = f_{(a, {\mathcal{L}}(g))}\left(
\frac{x}{1+(1-q)x} \right)\, ,
\end{eqnarray}
we easily obtain  from (\ref{IIIa}) that
\begin{eqnarray}
{\mathcal{D}}(A) = {\textsl{g}}^{-1}\left( \sum_k \varrho_k(q)
{\textsl{g}}({\mathcal{L}}_q \left(1/p_k\right)) \right) \, .
\label{VIb}
\end{eqnarray}
Moreover, if we set in the second part of axiom 3, $B = A$ then
${\mathcal{D}}(A)$ is given as
\begin{eqnarray}
{\mathcal{D}}(A) = f^{-1}\left( \sum_k \varrho_k(q)
f({\mathcal{L}}_q \left(1/p_k\right)) \right) \, . \label{IIIf}
\end{eqnarray}
Using the fact that two quasi-linear means with the same
${\mathcal{P}}$ are identical iff their respective
Kolmogorov--Nagumo functions are linearly related~\cite{Re2}, we
may write
\begin{eqnarray}
{\textsl{g}}(x) = f\left( \frac{-x + y}{1+(1-q)x}\right) =
\theta_q(y) f(x) \ + \ \vartheta_{q}(y)\, . \label{IIIc}
\end{eqnarray}
Here $y = {\mathcal{L}}_q(g)$. In order to solve (\ref{IIIc}) we
define $\varphi(x) = f(x) -f(0)$. With this notation
Eq.~(\ref{IIIc}) turns into
\begin{eqnarray}
\varphi\left(\frac{-x + y}{1+(1-q)x}\right) = \theta_q(y)
\varphi(x) \ + \ \varphi(y)\, , \; \; \;\; \; \varphi(0) = 0\, .
\label{IIId}
\end{eqnarray}
By setting $x = y$  we obtain that $\theta_q(y) = -1$, and hence
\begin{eqnarray}
\varphi(x + y + (1-q)xy) \ = \ \varphi(x) + \varphi(y)\, .
\label{IIIe}
\end{eqnarray}
According to axiom 1 we may now extend (\ref{IIIe}) to real valued
$x$ and $y$. Eq.(\ref{IIIe}) is Pixeder's functional equation
which can be solved by the standard method of
iterations~\cite{Acz1}. In~\cite{HLP1} we have shown that (\ref{IIIe})
has only one non-trivial class of solutions, namely
\begin{eqnarray}
\varphi(x) = \frac{1}{\alpha} \ln\left[  1+ (1-q)x \right]\, .
\label{IIIj}
\end{eqnarray}
$\alpha$ is here a free parameter. By inserting this solution back to
(\ref{IIIf}) we obtain
\begin{eqnarray}
{\mathcal{D}}_q(A) &=&   \frac{1}{1-q}\ \left( e^{-c(q)\sum_k
\varrho_k(q) \ln p_k}  -1 \right) = \frac{1}{1-q} \left( \prod_k
(p_k)^{-c(q)\varrho_k(q)} -1 \right)\, . \label{IIIg}
\end{eqnarray}
Note that the constant $\alpha$ got canceled. We have also
denoted the explicit order of the entropy ${\mathcal{D}}$ with the
subscript $q$. It remains to determine $c(q)$.  Utilizing the
conditional entropy constructed from (\ref{IIIg}) and using axiom
$3$, we obtain $c(q) = 1-q$. In result we can recast (\ref{IIIg})
into more expedient form. By utilizing
\begin{eqnarray}
\langle \ln {\mathcal{P}} \rangle_{q} \ = \ (1-q) \frac{\ud {\mathcal{I}}_q ({\mathcal{P}})}{\ud {q}} - {\mathcal{I}}_{q}({\mathcal{P}})\, .
\end{eqnarray}
the following results holds
\begin{eqnarray}
{\mathcal{D}}_q(A) \ = \ \frac{1}{1-q}\ \left( e^{-(1-q)^2
d{\mathcal{I}}_q/dq} \sum_{k=1}^n (p_k)^q -1 \right)\, .
\end{eqnarray}

%%%%%%%%%%%%%%%%%%%%%%%%%%%%%%%%%%%%%%%%%%%%%%%%%%%%%%%%%%%%%%
\subsection*{Restrictions on $q$ from the maximality axiom}
%%%%%%%%%%%%%%%%%%%%%%%%%%%%%%%%%%%%%%%%%%%%%%%%%%%%%%%%%%%%%%

In the foregoing proof we have used the axiom 2 to show that $\mathcal{L}(n+1) \geq \mathcal{L}(n)$, which in turn yielded $\mathcal{L}(n) = \ln_{\{q\}}(n)$, cf. Eq.~(\ref{III0a}). We have not, however, checked whether the global maximum is really at $\mathcal{P} = \{1/n,\dots,1/n\}$. In situation when
the entropy is a (Schur-)concave function on the probability space, we obtain the maximality directly. This is the case, e.g.,
for both R\'{e}nyi and THC entropy. Unfortunately, a (Schur-)concavity of $\mathcal{D}_q$ is ensured only for certain values of $q$ (as discussed in Section~\ref{sec: concavity}). Here we illustrate the fact that $\mathcal{D}_q$ can have maxima in other points than $\mathcal{P} = \{1/n,\dots,1/n\}$. To this end we note from (\ref{IIIh}) that because $\ln_{\{q\}}(x)$ is a monotonous function for $x>0$  and since $e^{-x}$ is a positive monotonous function on $\mathds{R}$, we can consider only  $\langle \ln \mathcal{P} \rangle_q$. For simplicity's sake, we present the analysis only for probability distribution of two events, i.e., $\mathcal{P} = \{p,1-p\}$. The analysis for more outcomes is similar, the only difference is that one has to employ
the Lagrange multipliers to account for the fact that the probability vector is confined on a simplex.

Stationary points of $\langle \ln \mathcal{P}  \rangle_q$ are solutions of the equation
\begin{eqnarray}
\mbox{\hspace{-10mm}}\frac{\ud}{\ud p} \left(\frac{p^q \ln p + (1-p)^q \ln (1-p)}{p^q+(1-p)^q}\right) &=&
%
%\begin{equation*}
%\frac{1}{(x^q+(1-x)^q)^2}\left[(x^q + (1-x)^q)\left(q x^{q-1} \ln x + x^{q-1} - q (1-x)^{q-1} \ln (1-x) - (1-x)^{q-1} \right) \right. -
%\end{equation*}
%\begin{equation*}
% - \left.(x^q \ln x + (1-x)^q \ln (1-x))(q x^{q-1} - q (1-x)^{q-1})\right]
% \end{equation*}
 \frac{1}{Z(q)^2}\left[p^{2q-1}-(1-p)^{2q-1}+p^{q-1}(1-p)^q-p^q(1-p)^{q-1}\right. \nonumber \\[2mm]
&&\left. -q p^q (1-p)^{q-1} \ln\left(\frac{1-p}{p}\right) +  q p^{q-1}(1-p)^q \ln\left(\frac{p}{1-p}\right)\right] \ = \ 0\, .
\label{A.22a}
\end{eqnarray}
The factor ${Z(q)^2} = {[p^q+(1-p)^q]^2}$ is positive and can be thus omitted from the further analysis. After the substitution $y={p}/{(1-p)}$
the previous equation
reduces to
\begin{equation}
y^{2q-1} - y^q(1-q \ln y) + y^{q-1}(1+q \ln y) -1 \ = \ 0\, ,
\end{equation}
or alternatively to
\begin{equation}
\Psi_q(y) \ \equiv \ q \ln y - \frac{1-y^{q-1} + y^q - y^{2q-1}}{y^q+y^{q-1}} \ = \ 0\, .
\end{equation}
The interesting property of $\Psi_q(y)$ is that $\Psi_q\left(\frac{1}{y}\right) = - \Psi_q(y)$ and $\Psi_1(y) = \ln y$.

The equation $\Psi_q(y) = 0$ has for $q \geq \frac{1}{2}$ only one solution, which is $y=1$, or equivalently $x= \frac{1}{2}$. However, for $q<\frac{1}{2}$, there occur two more solutions, related by the reciprocity relation. As a consequence, from the nature of $\langle \ln \mathcal{P} \rangle_q$ one can  deduce, that the point $x=\frac{1}{2}$ corresponds to the local minimum, while other two points represent global maxima. Eventually, the second axiom is violated for $q< \frac{1}{2}$ and  $\mathcal{D}_q$ is therefore well defined only for $q \geq \frac{1}{2}$.

%%%%%%%%%%%%%%%%%%%%%%%%%%%%%%%%%%%%%%%%%%%%%%%%%%%%%%%%%%%%%%%
\section{Basic properties of ${\mathcal{D}}_q$ entropy}\label{AppB}
%%%%%%%%%%%%%%%%%%%%%%%%%%%%%%%%%%%%%%%%%%%%%%%%%%%%%%%%%%%%%%%

In this appendix, we list some basic properties of the hybrid entropy ${\mathcal{D}}_q$.\\

Let us start with features that ${\mathcal{D}}_q$ shares with both R\'{e}nyi's and THC entropies. These are
\begin{tabbing}
~~~~~(a) ${\mathcal{D}}_q({\mathcal{P}} = \{ 1,0, \ldots,0 \}) =
0$\\
\\
~~~~~(b) ${\mathcal{D}}_q({\mathcal{P}}) \geq 0$\\
\\
~~~~~(c) ${\mathcal{D}}_1 = {\mathcal{I}}_1 = {\mathcal{S}}_1 =
{\mathcal{H}}$\\
\\
~~~~~(d) ${\mathcal{D}}_q$ involves a single free parameter - $q$\\
\\
~~~~~(e) ${\mathcal{D}}_q$ is symmetric, i.e.,
${\mathcal{D}}_q(p_1, \ldots, p_n) = {\mathcal{D}}_q(p_{k(1)},
\ldots, p_{k(n)})$\\
\\
~~~~~(f) ${\mathcal{D}}_q$ is bounded\\[-2mm]
\end{tabbing}

\noindent On the other hand, among features inherited from
R\'{e}nyi's entropy we can find that

\begin{tabbing}
~~~~~(g) ${\mathcal{D}}_q(A) = f^{-1}\left(\sum_k \varrho_k(q)
f({\mathcal{D}}_q(A_k))\right)$\\
\\
~~~~~(h) For single-dimensional statistical systems with
continuous PDF ${\mathcal{D}}_q(A)$ reduces to ${\mathcal{H}}$\\
\\
~~~~~(i) ${\mathcal{D}}_q$ is a strictly decreasing function of
$q$, i.e.,
$d{\mathcal{D}}_q/dq \leq 0$, for any $q > 0$
\end{tabbing}

\noindent Result (i) follows from the fact that ${\mathcal{D}}_q$
is a monotonically decreasing function of $A_q \equiv \sum_k
\varrho_k(q) \ln p_k$  (see Eq.(\ref{IIIg})) and that $A_q$ is a
monotonically increasing function of $q$, indeed
\begin{eqnarray}
\frac{dA_q}{dq} \ = \ \langle (\ln({\mathcal{P}}))^2 \rangle_q -
\langle \ln({\mathcal{P}}) \rangle_q^2  \ \geq \ 0 \, .
\label{IIIm}
\end{eqnarray}
Here $\langle \ldots \rangle_q$ is defined with respect to the
distribution $\varrho_k(q)$. The last relation in (\ref{IIIm}) is
Jensen's inequality. Note that $d{\mathcal{D}}_q/dq = 0$ happens
only for the degenerate case ${\mathcal{P}} = \{1, \ldots, 0\}$ (and ensuing permutations).\\
here
Finally, properties taken over from THC entropy include

\begin{tabbing}
~~~~~(j) $\max_{{\mathcal{P}}} {\mathcal{D}}_q({\mathcal{P}}) =
{\mathcal{D}}_{q}({\mathcal{P}} = \{ 1/n, \ldots, 1/n \}) = \ln_{\{q\}} n$ ~~~~~~~~(for $q>1/2$)\\
\\
~~~~~(k) ${\mathcal{D}}_q$ is $q$ non--extensive, i.e.,
${\mathcal{D}}(A\cup B) = {\mathcal{D}}(A) + {\mathcal{D}}(B|A) +
(1-q){\mathcal{D}}(A){\mathcal{D}}(B|A)$
\end{tabbing}

\section{Some essentials of the multifractal formalism}\label{AppC}
%%%%%%%%%%%%%%%%%%%%%%%%%%%%%%%%%%%%%%%%%%%%%%%%%%%%%%%%%%%%%%%%%

We present here some essentials of the fractal and multiftactal
calculus that are employed in the main body of the text.

%Let us begin with some fundamentals on fractals and multifractals.
{\em Fractals} are sets with a generally non--integer dimension
exhibiting property of self--similarity. The key characteristic of
fractals is a fractal dimension which is defined as follows:
Consider a set $M$ embedded in a $d$--dimensional space. Let us
cover the set with a mesh of $d$--dimensional cubes of size
$\varepsilon^d$ and let $N_{\varepsilon}(M)$ is a number of the
cubes needed for the covering. The fractal dimension of $M$ is
then defined as~\cite{Man1,Fed1}
\begin{equation}
D = - \lim_{\varepsilon \rightarrow 0} \frac{\ln
N_{\varepsilon}(M)}{\ln \varepsilon}\, . \label{B1}
\end{equation}
\noi  The dimension defined in (\ref{B1}) is also known as box-counting dimension.
In most cases of interest the latter
coincides with the Hausdorff--Besicovich dimension used by
Mandelbrot~\cite{Man1}.

{\em Multifractals}, on the other hand, are related to the study
of a distribution of physical or other quantities on a generic
support (be it or not fractal) and thus provide a move from the
geometry of sets as such to geometric properties of distributions.
Let us suppose that over some support (usually a subset of a
metric space) is distributed a probability of a certain
phenomenon. If we pave the support with a grid of spacing
$\varepsilon$ and denote the integrated probability in the $i$th
box as $p_i$, then the scaling exponent $a_i$ is
defined~\cite{Man1,Fed1}
\begin{equation} p_i (\varepsilon) \sim {\varepsilon}^{a_i}\, .
\label{B2}
\end{equation}
%
%\noindent so the density
%
%\begin{equation}
%\varrho_i = \frac{p_i}{l^d} \propto l^{\alpha_i -d}\, .
%\end{equation}
%
The exponent $a_i$ is called  singularity  or Lipshitz--H\"{o}lder
exponent. Counting boxes $N(a)$ where $p_i$ has $a_i \in (a,a +
da)$, the singularity spectrum $f(a)$ is defined
as~\cite{Man1,Fed1}
\begin{equation}
N(a) \sim {\varepsilon}^{-f(a)}\, . \label{B3}
\end{equation}
Thus a multifractal is the ensemble of intertwined (uni)fractals
each with its own fractal dimension $f(a_i)$. It is further
convenient to define a ``partition function"~\cite{Man1}
\begin{equation}
Z(q) = \sum_i (p_i(\varepsilon))^q = \int d a' \rho(a')
{\varepsilon}^{-f(a')} {\varepsilon}^{q a'}\, , \label{B5}
\end{equation}
($\rho(a)$ is a proportionality function having its origin in
relations (\ref{B2}) and (\ref{B3})). In the small $\varepsilon$
limit the method of steepest descent yields the scaling
\begin{equation}
Z(q)\sim {\varepsilon}^{\tau(q)}\, , \label{tau}
\end{equation}
with
\begin{equation}
\tau(q) = \min_{a} [q a - f(a)], \,\, f'(a) =q  \,\,\,
\mbox{and}\,\,\, \tau'(q) = a(q)\, . \label{B4}
\end{equation}
These are precisely Legendre transform relations. Scaling exponent
$\tau$ is often called the correlation exponent. Legendre
transform (\ref{B4}) ensures that pairs $f(a), a$ and $\tau(q),
q$, are conjugates comprising the same mathematical content.

%Connection of R\'{e}nyi entropies with multifractals is frequently
%introduced via generalized dimensions
%%
%\begin{eqnarray}
%D_q = \lim_{l\rightarrow 0} \left( \frac{1}{(q-1)} \frac{\log
%Z(q)}{\log l} \right) = -\lim_{l \rightarrow 0} {\cal{I}}_q
%(l)/\log_2 l\, . \label{gendim1}
%\end{eqnarray}
%%
It is an important consequence of (\ref{B5}) that the relative fluctuations of the Lipshitz--H\"{o}lder exponent $a$
around its mean value $\langle a \rangle_q$ are
very small in the $\varepsilon \rightarrow 0$ limit. This is because
\begin{eqnarray}
&&\partial^2(\ln Z(q))\partial q^2 \ = \ \left[\langle a^2 \rangle_q \ - \ \langle a \rangle_q^2\right] (\ln \varepsilon)^2 \ \equiv \ (\Delta a)^2 (\ln \varepsilon)^2 \, ,\label{C7a} \\[2mm]
&&\partial^2(\tau \ln \varepsilon)\partial q^2 \ = \ \left({da}/{dq}\right)\ln \varepsilon \ = \
\left[{1}/{f''(a)} \right] \ln \varepsilon \, .  \label{C7b}
\end{eqnarray}
Since both left-hand sides in (\ref{C7a}) and (\ref{C7b}) are identical, we  can infer from a finiteness of $f''(a)$ that
the standard deviation of $a$ is of order $1/\sqrt{-\ln \varepsilon}$.
So for small $\varepsilon$ the $a$-fluctuations become negligible and almost all $a$
equal to $\langle q \rangle_q$. Note also that because the variance $(\Delta a)^2 >0 $  and $\ln \varepsilon <0$, we have that $f''(a) <0$, i.e., the $f(a)$ function is concave.

The fact that for a given $q$ the total probability of a phenomenon with a scaling exponent $a_i$ is concentrated around the value $a_i \sim \langle a\rangle_q$  is known as the curdling theorem~\cite{Man1} (or Billingsley theorem~\cite{Bill})
and it represents a particular example of the so-called measure concentration
phenomenon~\cite{JA4}.

Multifractal formalism has direct applications in the turbulent
flow of fluids~\cite{ArAr1}, percolation~\cite{Ah1},
diffusion--limited aggregation (DLA) systems~\cite{NST1}, DNA
sequences~\cite{ZGU1}, finance~\cite{VVA1,JK}, string
theory~\cite{MMW1}, etc.. In chaotic dynamical systems all
${\mathcal{I}}_q$ are necessary to describe uniquely, e.g.,
strange attractors~\cite{HaPr1}. More generally, one may
argue~\cite{PJ1} that when the outcome space is discrete then all
${\mathcal{I}}_q$ (or ${\mathcal{S}}_q$) with $q \in [1, \infty)$ but
are needed to reconstruct the underlying distribution, while when
the outcome space is $d$-dimensional subset of ${\mathbb{R}}^d$
then all ${\mathcal{I}}_q$ (or ${\mathcal{S}}_q$), $q \in (0,
\infty)$, are required to pinpoint uniquely the underlying PDF.
The latter can be viewed as the information--theoretic variants of
Hausforff's moment problem of mathematical statistics.

The connection of R\'{e}nyi entropies with multifractals is
established via relation (\ref{B5}). Note particularly that when
$\varepsilon$ is finite then ${\mathcal{I}}_q$ plays the r\^{o}le
of the Helmholtz free energy.  Closer analysis of the related
implications can be found, e.g., in Refs.~\cite{PJ1,PJ2}.

%%%%%%%%%%%%%%%%%%%%%%%%%%%%%%%%%%%%%%%%%%%%%%%%%%%%%%%%%%%%%%%%%%%%%%%%
%\section*{References}
%%%%%%%%%%%%%%%%%%%%%%%%%%%%%%%%%%%%%%%%%%%%%%%%%%%%%%%%%%%%%%%%%%%%%%%

\end{document}